\documentclass[twocolumn,final]{svjour3}
\usepackage{amssymb}
\usepackage{amsmath}
\usepackage{graphicx}
\usepackage{mathtools}	
\usepackage{multirow}
\usepackage{subfig}
\usepackage{siunitx}
\usepackage{gensymb}
\usepackage[draft]{pgf}
\usepackage{algorithm}
\usepackage[noend]{algpseudocode}
\usepackage[utf8]{inputenc}
\usepackage{booktabs}
\usepackage{mwe}
\usepackage{url}
\graphicspath{{Images/}}
\urldef{\mailsc}\path|matt.dale@york.ac.uk|    

\newif\ifisProcRoySocA
\isProcRoySocAfalse

\usepackage[colorinlistoftodos]{todonotes}

{}% for the vertical padding
\def\makeheadbox{\relax}
\date{}

\begin{document}
\raggedbottom

\title{A Substrate-Independent Framework to Characterise Reservoir Computers}

\titlerunning{}

\author{Matthew Dale$^{1,3}$, Julian F. Miller$^{3}$, Susan Stepney$^{1,3}$, Martin A. Trefzer$^{2,3}$}

\authorrunning{Matthew Dale, Julian F. Miller, Susan Stepney, Martin A. Trefzer}
\institute{
$^1$Department of Computer Science, University of York, UK \\
$^2$Department of Electronic Engineering, University of York, UK  \\
$^3$York Cross-disciplinary Centre for Systems Analysis \\
% * <susan.stepney@york.ac.uk> 2018-05-25T07:56:45.763Z:
%
% ^.
\url{matt.dale@york.ac.uk}}

\maketitle

\begin{abstract}

The Reservoir Computing (RC) framework states that any non-linear, input-driven dynamical system (the \textit{reservoir}) exhibiting properties such as a fading memory and input separability can be trained to perform computational tasks. This broad inclusion of systems has led to many new physical substrates for RC. Properties essential for reservoirs to compute are tuned through reconfiguration of the substrate, such as change in virtual topology or physical morphology. As a result, each substrate possesses a unique ``quality'' -- obtained through reconfiguration -- to realise different reservoirs for different tasks. Here we describe an experimental framework to characterise the quality of potentially \textit{any} substrate for RC. Our framework reveals that a definition of quality is not only useful to compare substrates, but can help map the non-trivial relationship between properties and task performance. In the wider context, the framework offers a greater understanding as to what makes a dynamical system compute, helping improve the design of future substrates for RC.


\end{abstract}

% quick recap on RC framework
 \section{Introduction}

Reservoir Computing (RC) first emerged as an alternative method for constructing and training recurrent neural networks~\cite{schrauwen2007overview,verstraeten2007experimental}. The method primarily involved constructing a random fixed recurrent network of neurons, and training only a single linear readout layer. It was found that random networks constructed with certain dynamical traits could produce state-of-the-art performance without the laborious process of training individual internal connections. The concept later expanded to encompass any high dimensional, input-drive\-n dynamical system that could operate within specific dynamical regimes, leading to an explosion in new reservoir computing substrates\footnote{The term ``substrate'' is used here to refer to any physical or virtual system that realises a reservoir computer: any dynamical system featuring configurable parameters and a method to observe system states.}.

The ability to perform useful information processing is an almost universal characteristic of dynamical systems, provided a fading memory and linearly independent internal variables are present \cite{dambre2012information}. However, each dynamical system will tend to suit different tasks, with only some performing well across a range of tasks.

In recent years, reservoir computing has been applied to a variety of physical systems such as optoelectronic and photonic~\cite{appeltant2011information,vandoorne2011parallel}, quantum~\cite{fujii2017harnessing,obst2013nano,torrejon2017neuromorphic}, disordered and self-organising~\cite{dale2016evolving,stieg2012emergent}, magnetic~\cite{jensen2018computation,prychynenko2018magnetic}, and memristor-based~\cite{du2017reservoir} computing systems. The way in which each substrate realises a reservoir computer varies. However, each tends to implement, physically or virtually, a network of coupled processing units. 

In each implementation, the concept is to utilise and exploit the underlying physics of the substrate, to embrace intrinsic properties that can improve performance, efficiency and/or computational power. Each substrate is configured, controlled and tuned to perform a desired functionality, typically requiring the careful tuning of parameters in order to produce a working, or optimal, physical reservoir computer for \textit{ad hoc} problems.

Despite the recent advances of new physical reservoir systems, basic questions for reservoir computing are still unanswered. These open problems are summarised and explained in~\cite{goudarzi2016reservoir}. Relevant questions include: What class of problems can RC solve efficiently? What is the role of heterogeneous structure in RC? What are the limits and benefits of a given physical system for RC? What are the benefits of a physical implementation? To answer these questions, and for the field to move forward, a greater understanding is required about the computational expressiveness of reservoirs and the substrates they are implemented on, if not to at least determine what tasks, for what substrates, are realistically solvable.

In the terminology used here, a \textit{reservoir} represents the resulting abstract system and its dynamics instantiated by (typically, but not limited to) a single, typically static, configuration of the substrate. For an artificial recurrent neural network, implemented \textit{in silico}, configuration may refer to a set of trained connection weights, defined neuron types and topology. For another substrate, configuration may refer to the physical morphology, physical state, external control signals, or complexification of the driving input signal. 
The number of possible reservoir systems realisable by a substrate depends upon the number of free parameters, and the distinct dynamical behaviours resulting from those parameters. For unconstrained substrates, limited only by the laws of physics, this number may be vast. Yet this does not imply that every such configuration and corresponding reservoir is practical or useful. This also does not imply that each new configuration leads to a different reservoir system; the same or similar dynamical behaviour may be produced by different configurations. The mapping between substrate configuration and instantiated reservoir may be complex.

Here we present a practical framework to measure the computational expressiveness of physical or virtual substrates, providing a method to characterise and measure the reservoir computing \textit{quality} of substrates. 

A higher quality substrate is one that can realise more \textit{distinct} reservoirs through configuration, giving it greater expressiveness and higher dynamical freedom, and so a greater \textit{capacity} to tackle very different tasks. Quality is quantified and measured here as the number of distinct reservoirs, or dynamical behaviours, a single substrate can exhibit. The number of  reservoirs, rather than configurations, is what is important. This does not imply that substrates with fewer available configuration degrees of freedom perform poorly at every task;
they may perform very well at specific tasks within their dynamical range, but are likely to perform poorly when evaluated across a broad range of tasks. 

%%%%%%%%%%%%%%%%%%%%%%%%%%% Figure %%%%%%%%%%%%%%%%%%%%%%%%%%%
\begin{figure*}[tp]
\centering
\includegraphics[width=0.75\textwidth,trim=4cm 1cm 3.5cm 0cm,clip]{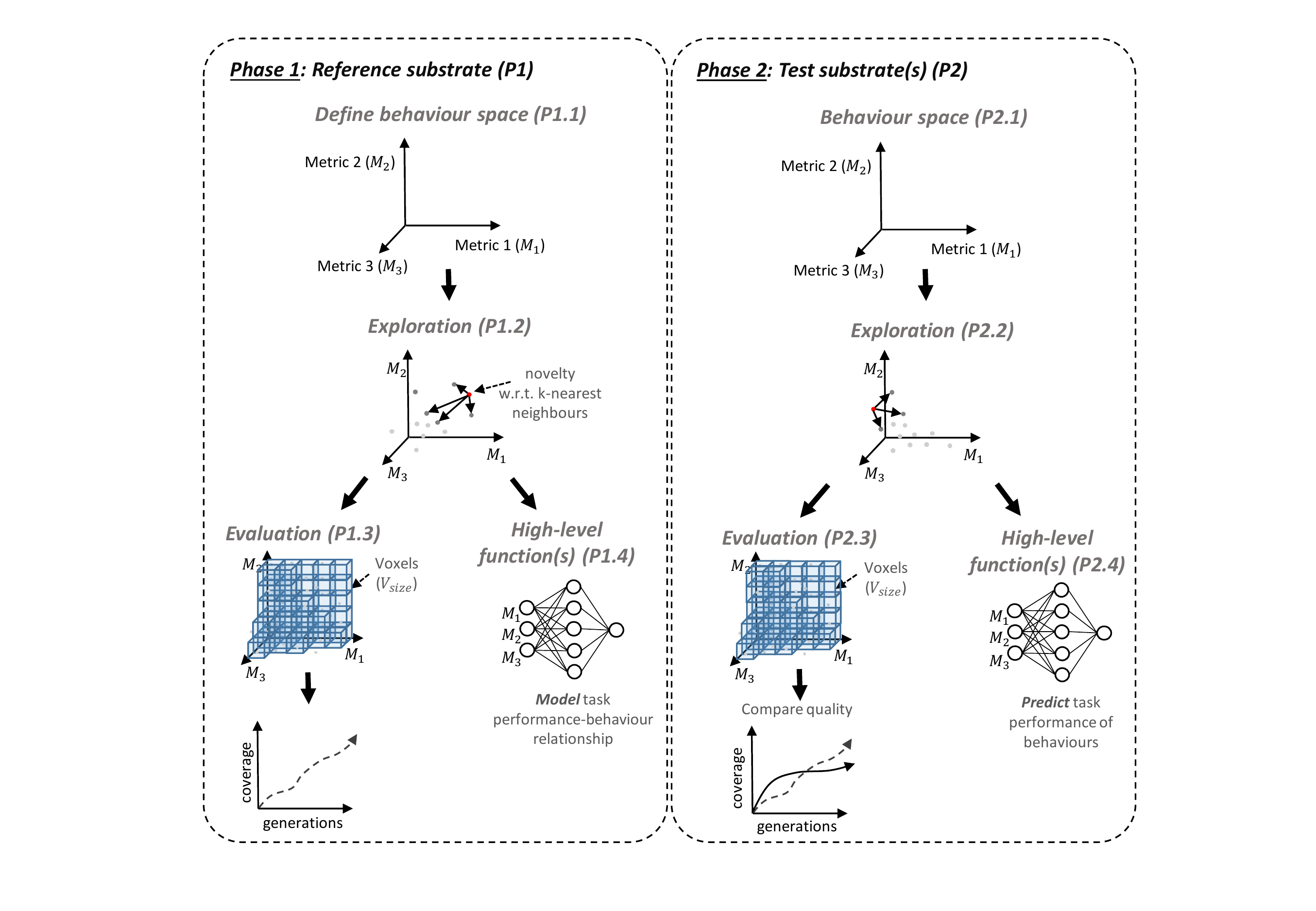}
\caption{Framework phases and levels.}
\label{fig: bblocks}
\end{figure*}
%%%%%%%%%%%%%%%%%%%%%%%%%%% Figure %%%%%%%%%%%%%%%%%%%%%%%%%%%

To characterise the quality of different substrates, we present the CHARC (CHAracterisation of Reservoir Computers) framework.
The framework has a basic underlying structure, which can be extended if needed. To demonstrate the framework, it is applied to three different substrates: Echo State Networks (ESN)~\cite{jaeger2001echo}, 
simulated Delay-based Reservoirs (DR)~\cite{appeltant2011information,ortin2015unified},
and a physical carbon nanotube composite~\cite{dale2016evolving}. 
The definitions, techniques and substrate-independence of the framework are evaluated using a number of common benchmark tasks. 

The rest of the paper describes the framework and the techniques used within it, beginning with a description of the workflow, the task-independent properties and search procedure used to characterise the substrate. 

%--------------------------------------------------------
\section{Framework Outline}
\label{sec:outline}

The basic structure and flow of the framework is presented in Fig.~\ref{fig: bblocks}. The complete  characterisation process is divided into a series of phases and levels. 
In phase one ($P1$), a reference substrate is evaluated, forming the basis against which to compare quality values.
%To make good use of the framework, and to assess its accuracy and validity, the first phase  requires  
In phase two ($P2$), the test substrate is assessed and compared to the reference.

\subsection{Basic levels} \label{sec:basic}

The three basic levels required for each phase are \textit{definition}, \textit{exploration}, and \textit{evaluation}. Additional levels may be added, providing further functions that can be used to manipulate, model and learn from the data produced by the characterisation process. Here, an additional level is used to validate and determine the reliability and substrate-independence of the overall  framework; others are also possible, see section~\ref{sec: High level functions}.  

In general, each level requires the results from the previous level. Techniques applied at each level are flexible, and may be substituted with alternative approaches. The techniques and measures used here are simple, and provide a good foundation to demonstrate the framework's concept.

\begin{figure}[tp]
\vspace{0.5cm}
\centering
\includegraphics[width=0.4\textwidth]{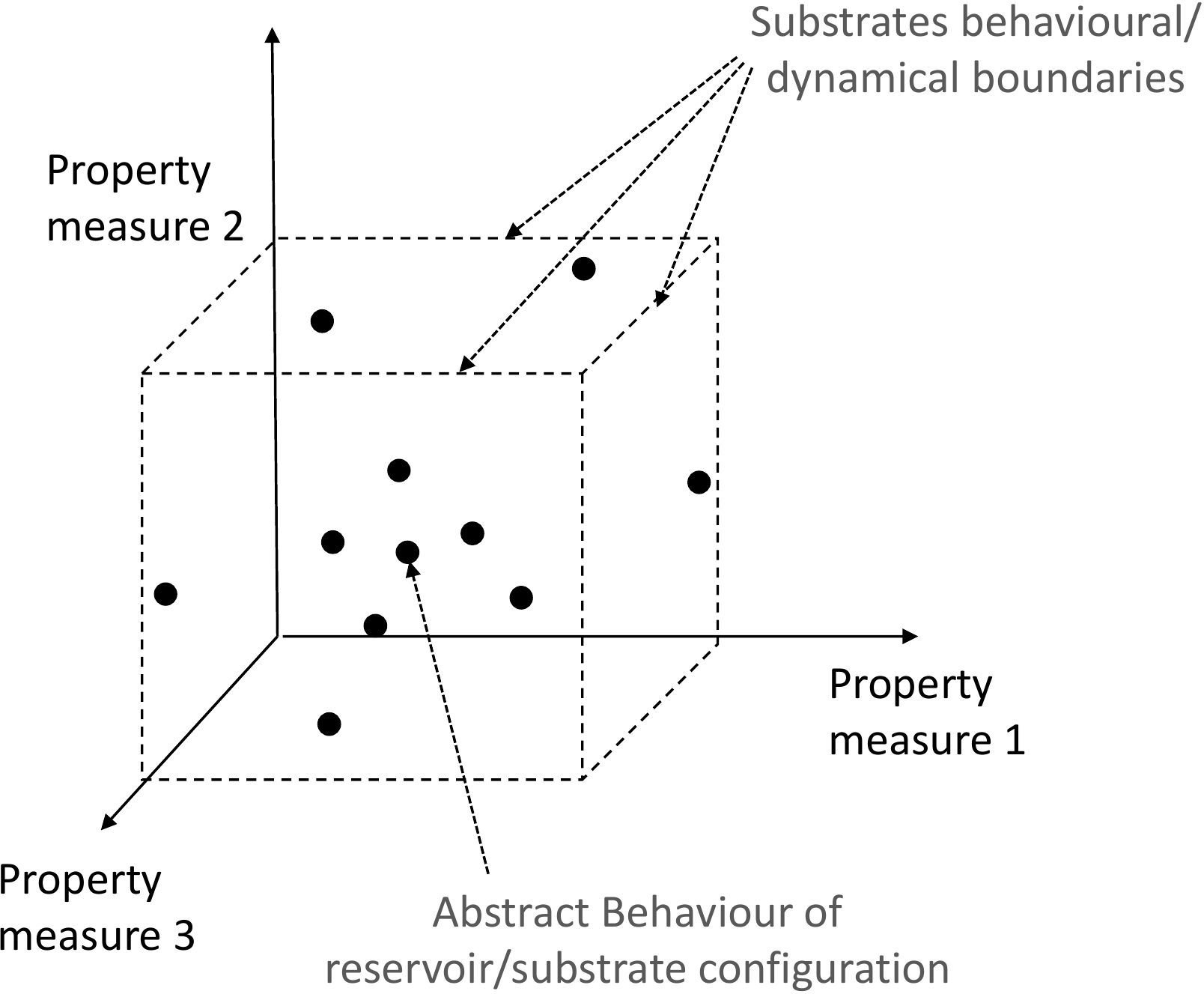}
\caption{Example of a 3-Dimensional Behaviour Space. Here each abstract behaviour is relative to the three chosen property measures. Given enough time to explore the space, the substrate's dynamical/behavioural limitations become apparent. }
\label{fig: example behaviour space}
\end{figure}

The \textit{definition} level ($P1.1$, $P2.1$) defines the reservoir \textit{behaviour} space to be explored. The behaviour space represents the abstract behaviour of the configured substrate relative to measures of dynamical properties, and is the space in which \textit{quality} is measured. 
%At this point, it is important to emphasise that the exploration procedure is not conducted in the configuration parameter space, but the space of possible reservoirs behaviours. 
The framework uses $n$ measures (see example in Fig.~\ref{fig: example behaviour space})
to define the axes of the \textit{n}-dimensional behaviour space. 
See section~\ref{sec: metrics} for the measures used here.

The \textit{exploration} level ($P1.2$, $P2.2$) measures the quality, by determining how much of the behaviour space is realisable through substrate configurations. 
%To determine a true measure of quality, rather than an approximation, would require 
An exhaustive search of the substrate's parameter space is infeasible. Instead, the use of diversity search algorithms~\cite{pugh2016quality} is recommended. 
These exploration techniques, based on evolutionary algorithms, can characterise the behaviour range and dynamical freedom of the substrate. 
 
The \textit{evaluation} level ($P1.3$, $P2.3$) estimates quality, by using the behaviours discovered from the exploration level. 
%As quality is defined as the total number of distinct reservoirs/dynamical behaviours the substrate can exhibit when reconfigured, we measure the spread and number of behaviours discovered. 
The behaviour space is divided into discrete voxels; the total number of voxels occupied by discovered behaviours provides the final quality value of the substrate. 
In $P2.3$, the quality of the test substrate is compared with the quality of the reference substrate from $P1.3$.  
%%%%%%%%%%%%%%%%%%%%%%%%%%% Figure %%%%%%%%%%%%%%%%%%%%%%%%%%%
%%%%%%%%%%%%%%%%%%%%%%%%%%% Figure %%%%%%%%%%%%%%%%%%%%%%%%%%%
\subsection{Additional levels providing further functions} \label{sec: High level functions}

Additional levels can be added to the framework to extract further features about the substrate and the behaviour space representation. They need not necessarily relate to the evaluation level (the quality value), and may work independently of it. Example additional levels include: modelling the relationships between the behaviour space and task-specific performances; modelling the relationships between the behaviour space and configuration space. Such relationships can be modelled and learnt using machine learning techniques. 

Here, one additional level is created: a \textit{learning} level ($P1.4$, $P2.4$). 
The learning level is used here to evaluate whether the framework is reliable (that the behaviour metrics capture the underlying reservoir properties) and substrate-independent (that behaviours lear\-ned in one substrate can be transferred to a different substrate). To achieve independence, the reliability of the behaviour space representation should be high. In reality, due to noise and translation between the simulated and physical domain, we require reliability above some acceptable threshold.

Further levels building on the exploration and learning levels are also possible. For example, the discovered behaviours can provide a reduced search space from which to rank and find optimal reservoirs for a particular task. As the number of tasks increases, this reduced search space decreases the required time to find good task-specific reservoirs without having to repeatedly search over the full configuration space.

% -----------------------------------------------------------
\subsection{Task-Independent Properties} \label{sec: metrics}

In order to form a suitable behaviour space, we need to define each dimension of the space carefully. 
Some potentially interesting properties are difficult, if not impossible, to transfer across all substrates. For example, measures that require access to the system's internal workings will not transfer to black-box systems.
Measures used with the framework must represent only the observable behaviour of the system, independent of its implementation.

In general, the behaviour space should be created using as many uncorrelated measures as possible, representing different computational and dynamical properties. This will improve the reliability of the framework, but result in a larger space to explore, requiring more evaluations to build a useful characterisation.%by creating a more faithful behavioural representation of the system. 

In the work here, three common property measures are taken from the reservoir computing literature to form the behaviour space. These measures are the Kernel Rank, Generalisation Rank, and Memory Capacity. 

\textit{Kernel Rank} (KR) is a measure of the reservoir's ability to separate distinct input patterns \cite{legenstein2007edge}. 
It measures a reservoir's ability to produce a rich non-linear representation of the input \(u\) and its history $u(t-1), u(t-2), \ldots$. This is closely linked to the \textit{linear separation property}, measuring how different input signals map onto different reservoir states. As many practical tasks are linearly inseparable, reservoirs typically require some non-linear transformation of the input. KR is a measure of the complexity and diversity of these non-linear operations performed by the reservoir.

\textit{Generalisation Rank} (GR) is a measure of the reservoir's capability to generalise given similar input streams \cite{legenstein2007edge}. 
It attempts to quantify the generalisation capability of a learning device in terms of its estimated VC-dimension~\cite{vapnik1999overview}, i.e., how well the learned non-linear function generalises to new inputs. In general, a \textit{low} generalisation rank symbolises a robust ability to map similar inputs to similar reservoir states, rather than overfitting noise.

Reservoirs in ordered dynamical regimes typically have low ranking values in both KR and GR, and in chaotic regimes have both high. A rule-of-thumb is that good reservoirs  possess a high KR and a low GR~\cite{busing2010connectivity}. In terms of matching reservoir dynamics to tasks, the precise balance will vary. 

A unique trait that physical and unconventional substrates likely possess is the ability to feature multiple time-scales and possess tunable time-scales through reconfiguration, unlike their more conventional reservoir counterparts.

Another important property for reservoir computing is memory, as reservoirs are typically configured to solve temporal problems. (A substrate without memory may still be computationally interesting for solving non-temporal problems.)
A simple measure for reservoir memory is the \textit{linear short-term memory capacity} (MC). This was first outlined in~\cite{jaeger2001short} to quantify the echo state property. For the echo state property to hold, the dynamics of the input driven reservoir must asymptotically wash out any information resulting from initial conditions. This property therefore implies a fading memory exists, characterised by the short-term memory capacity.

A full understanding of a reservoir's memory capacity, however, cannot be encapsulated through a linear memory measure alone, as a reservoir will possess some non-linear memory. Other memory measures proposed in the literature quantify other aspects of memory, such as the quadratic and cross-memory capacities, and total memory of reservoirs using the Fisher Memory Curve~\cite{dambre2012information,ganguli2008memory}. The linear measure is used here to demonstrate the framework; additional measures can be added as needed.

% -------------------------------------------
\subsection{Behaviour Space Exploration}

To characterise the reservoir behaviour space, the search must explore without optimising towards any particular  property values. 
%The optimisation of the measures is often unnecessary, time-consuming, and may in fact hinder performance. 
A balance between properties is essential to match reservoir dynamics to tasks. However, determining the right balance is challenging. During the characterisation process, the exact balance required for specific tasks is irrelevant. Instead, the focus is to explore and map the space of possible trade-offs the substrate can exhibit, and use this to determine substrate quality.

For the framework to function, the mapped reservoir behaviour space requires substrate-independence,
so the exploration cannot be conducted, or measured, in the substrate-specific parameter space.
Also, the exploration must be able to function without  prior knowledge of how to construct reservoirs far apart from each other in the behaviour space,
as diversity in observed dynamics is not easily related to diversity in substrate-specific parameters.
 
Here, exploration is performed using the open-ended Novelty Search (NS) algorithm \cite{lehman2008exploiting,lehman2010efficiently,lehman2011abandoning}, one of several possible diversity algorithms~\cite{pugh2016quality}. 
%Diversity algorithms have been shown to counter-intuitively outperform objective-based methods in problem domains that have deceptive fitness landscapes~\cite{risi2009novelty}. This advantage exists because the solution space's are typically sparse, or difficult to navigate due to many local optima. 
%These algorithms overcome deceptive spaces through balancing global exploration and exploiting local niches.
Novelty Search increases the selection pressure of an underlying evolutionary algorithm towards novel behaviours far apart in the behaviour space. The full details of our Novelty Search implementation are given in the Appendix~\ref{app: novelty search}.

%Given enough time, the process will also provide an approximation of the dynamical boundaries of the substrate (see Fig.~\ref{fig: example behaviour space}). This can then be used determine the practical use, if any, of the substrate, or whether the selected method of configuration and observation (which may itself be improved) is appropriate. 

% -----------------------------------------------------------
\section{Phase One: Reference Substrate}
\label{sec:phaseone}

Phase one establishes a suitable \textit{reference} substrate to compare against. Here, we use Recurrent Neural Networks (RNN) that closely resemble Echo State Networks (ESN)~\cite{jaeger2001short} as the reference. These are well -established state-of-the-art reservoir ``substrates''. RNNs are flexible, universal approximators of dynamical systems~\cite{funahashi1993approximation} producing a vast range of dynamics when reconfigured.
 
For a standard ESN, the reservoir state update equation \(x(t)\) is:
\begin{equation} \label{eq: esn state update}
x(t) = f(W_{in}u(t) + Wx(t-1) + W_{fb}y(t))
\end{equation}
where $f$ is the neuron activation function (typically a sigmoid) and the weight matrices \(W_{in}\), \(W\), and \(W_{fb}\) are matrices of connection weights to inputs (\(W_{in}\)), internal neurons (\(W\)), and from the output to internal neurons (\(W_{fb}\)). In many cases, the feedback weights \(W_{fb}\) are unused and the other weight matrices are selected from a random distribution, then scaled globally. 

The final trained output \(y(t)\) is given when the reservoir states \(x(t)\) are combined with the trained readout layer \(W_{out}\): 
\begin{equation} \label{eq: esn output}
y(t)= W_{out}x(t)
\end{equation}

Training of the readout is typically carried out in a supervised way using one-shot linear regression with a teacher signal. A practical guide to creating and training ESNs is given in~\cite{lukovsevivcius2012practical}.

\subsection{Demonstrating and validating the framework} \label{sec:eval}
In a typical use of the framework, one would now perform the various levels of Phase one to characterise the ESN reference substrate.
Here we do more, performing several experiments to demonstrate why certain choices have been made, to explore the the framework in action, and to determine the reliability of the results.

Here, four sizes of ESNs are used for the purpose of demonstrating the framework. The four network sizes chosen have 25, 50, 100, and 200 nodes. This small spectrum from simple to complex reservoirs provides a useful test suite. Each size is a constrained version of the general ESN substrate, and exhibits different ranges of dynamical behaviours. 

% BS of all spaces
%%%%%%%%%%%%%%%%%%%%%%%%%%% Figure %%%%%%%%%%%%%%%%%%%%%%%%%%%
\begin{figure*}[tt]
\centering
\subfloat[50 node]{\includegraphics[width=0.47\textwidth,trim=2cm 7cm 2cm 7cm,clip]{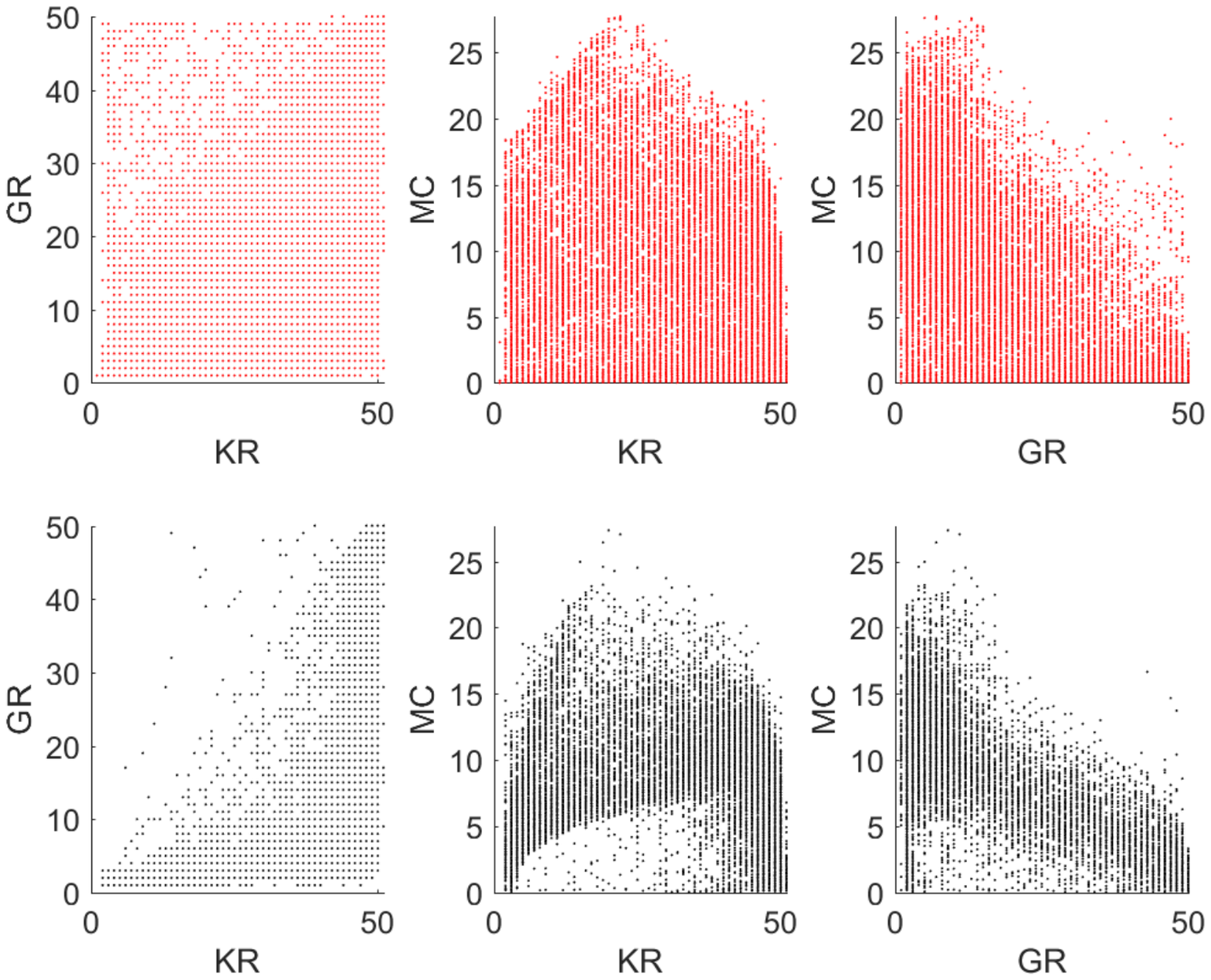}\label{fig: 25coverage of rand}} \hspace{0.5cm}
\subfloat[200 node]{\includegraphics[width=0.47\textwidth,trim=2cm 7cm 2cm 7cm,clip]{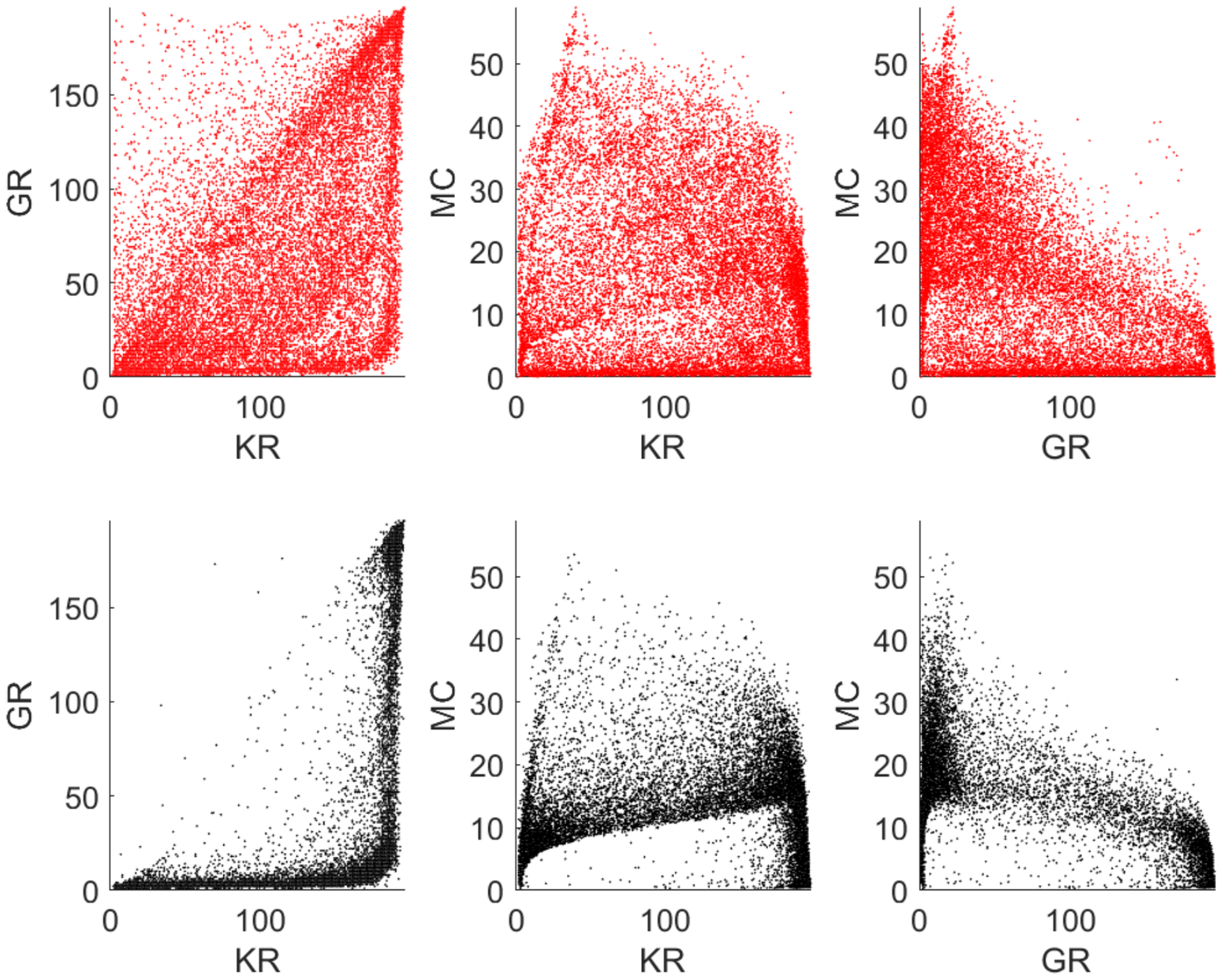}\label{fig: 200coverage of NS}}

\caption{Behaviours discovered using Novelty Search (red, top row) and random search (black, bottom row) for network sizes of: a) 50 nodes, b) 200 nodes. The 3D behaviour space is shown as three projections.  \label{fig: NS vs rand BS}}
\end{figure*}
%%%%%%%%%%%%%%%%%%%%%%%%%%% Figure %%%%%%%%%%%%%%%%%%%%%%%%%%%

\subsection{Novelty vs. Random Search} \label{sec: ns vs rand}
Here we apply the exploration process $P1.2$, and evaluate the use of Novelty Search (NS) by comparing it to random search, determining its usefulness for characterising a substrate. If NS performs well, if it discovers a greater variation in behaviours than random search within the same time, across network sizes, we argue it will continue to be advantageous for different substrates. 

First, we compare Novelty Search and random search visually. The hypothesis here is that Novelty Search can cover a greater area of the behaviour space within the same number of search evaluations. 

The results of this experiment show that for every network size, Novelty Search expands further in all dimensions of the behaviour space. In Fig.~\ref{fig: NS vs rand BS}, the explored spaces of the 50 and 200 node ESNs using both search techniques are plotted. In total, approximately 20,000 configurations from 10 separate runs are displayed.  

Random search (in black), which selects weights and scaling parameters from uniform random distributions, appears to produce similar patterns in the behaviour space across all network sizes. These patterns show spar\-se regions that are difficult to discover, and dense areas that are frequently visited despite different configuration parameters. As network size increases, random sear\-ch tends to find it more challenging to uniformly cover the behaviour space, suggesting it becomes less effective as substrate complexity increases.

Novelty search (in red) covers the behaviour space more uniformly, filling sparse regions and expanding into areas undiscovered by the random search. It does this within the same number of network evaluations, showing itself to be a more effective search technique than simply sampling the configuration space from a random uniform distribution. 

\subsection{Quality Measure} \label{sec: quality measure}
Here we perform the evaluation process $P1.3$ on the behaviours discovered by NS above, in order to evaluate the voxel-based quality measure proposed to quantify the coverage of the behaviour space, and thus quality. 

To measure quality and coverage of the behaviour space, standard statistical measures of dispersion such as standard deviation, mean absolute deviation, and inter-quartile range are not suitable by themselves: they downplay outliers, whereas the aim is to measure both the volume and the boundaries of the region explored. For this reason, a voxel-based measure is adopted here.
Discovered behaviour instances occupying the same vox\-el are counted once, thereby grouping similarly behaved reservoirs as a single behaviour voxel. 

% In our three-dimensional example, this discretised behaviour space is captured by a large cube encompassing the maximum possible range of behaviours, partitioned into smaller voxel cubes. The smallest possible voxel size is $1\times1\times1$: the smallest discretised value of the KR and GR property measures. 

In our three-dimensional example, the discovered behaviours define the bounds of the measurable behaviour space; a large cube. The space is then discretised and partitioned into smaller voxel cubes. The smallest possible voxel size is $1\times1\times1$: the smallest discretised value of the KR and GR property measures. 

Voxel size needs to be chosen carefully in order to accurately compare substrates.
If the voxel size is too small, every occupied voxel will contain exactly one explored reservoir behaviour, and the quality measure will merely record the number of search points evaluated.
If the voxel size is too large, the quality measure will be too coarse grained to make distinctions.

%The coverages of each search method and network size using this minimal voxel size are given in Fig.~\ref{fig: 1x1x1 voxel}. There we see that small voxel sizes significantly overestimate the dynamical freedom/quality of networks explored using novelty search. Smaller networks such as the 25 and 50 node ESNs are seen to occupy similar or more voxels than larger networks, suggesting similar or better quality. However, when visually comparing each explored behaviour space (see Fig.~\ref{fig: ESN BS overlay}) we see the measure fails to account for diversity and spread of behaviours. This demonstrates the importance of selecting an appropriate voxel size, as a voxel size too small can not differentiate between local areas that are highly populated, and fewer behaviours spread across a greater distance. 

%To reduce the problem, the voxel size must be increased. By how much depends on the size of the behaviour spaces being compared. As a guide, when comparing drastically different systems, a larger voxel size will tend to differentiate better. Of course, a voxel size too big will also struggle to differentiate between systems. Because of this potential biasing problem, a visual inspection of the behaviour space is always recommended. Examples of the measure conducted using different voxel sizes are given in Fig.~\ref{fig: voxel measures}. 

Experiments to investigate the effect of voxel size are given in appendix~\ref{app:voxel}.
These lead us to choose a voxel size of $V_{size} = 10\times10\times10$ for the rest of this paper. 

The quality of a tested substrate is equal to the final coverage value. As voxel size and total number of evaluations both affect this value, the reference and test substrate should be compared using the same framework parameters.

\subsection{Reliability of the Behaviour Space} \label{sec: accuracy of BS}

In the last part of phase one addressed here, $P1.4$, the reliability of the behaviour space is measured, to demonstrate that the framework produces usable results. The outcome of this measure is also used to determine that the behaviour space is independent of the substrate implementation, $P2.4$, section~\ref{sec: pred across subs}. If the reliability is poor, independence is difficult to measure and interpret.

To assess reliability and independence, concepts such as the \textit{representation} relation and commuting diagrams from Abstraction/Representation (A/R) theory~\cite{horsman2014does} are adapted to form a testable hypothesis. In A/R theory, a framework is proposed to define when a physical system computes. Using those concepts, one can assess whether an abstract computational model reliably represents computation performed by a physical system.

Our hypothesis for the framework is that if the abstract reservoir space is truly representative of system dynamics, and independent of its implementation, it should hold that similar behaviours across substrates produce similar task performances. This hypothesis was conceived using A/R commuting diagrams as a template, where if the computational model faithfully represents the computation of the physical system, one can predict how the physical system states will evolve. 

To test the hypothesis, first the relationship between task performance and reservoir behaviour is modelled. The reliability of this model, measured as the prediction error of the model, indicates how well the behaviour space captures the computation occurring within the substrate.

As explained in~\cite{goudarzi2016reservoir}, relating property measures to expected performance across many tasks is a non-trivial problem, as good properties for one task are often detrimental to another. Therefore, no single set of measured values will lead to high performance across all tasks. However, the relationship between behaviour measure values and a single task are often simpler to determine; these are the relationships to be modelled.

%%%%%%%%%%%%%%%%%%%%%%%%%%% Figure %%%%%%%%%%%%%%%%%%%%%%%%%%%
\begin{figure*}[tp]
\centering
\includegraphics[width=0.6\textwidth,trim=1.25cm 7.5cm 1.25cm 8cm,clip]{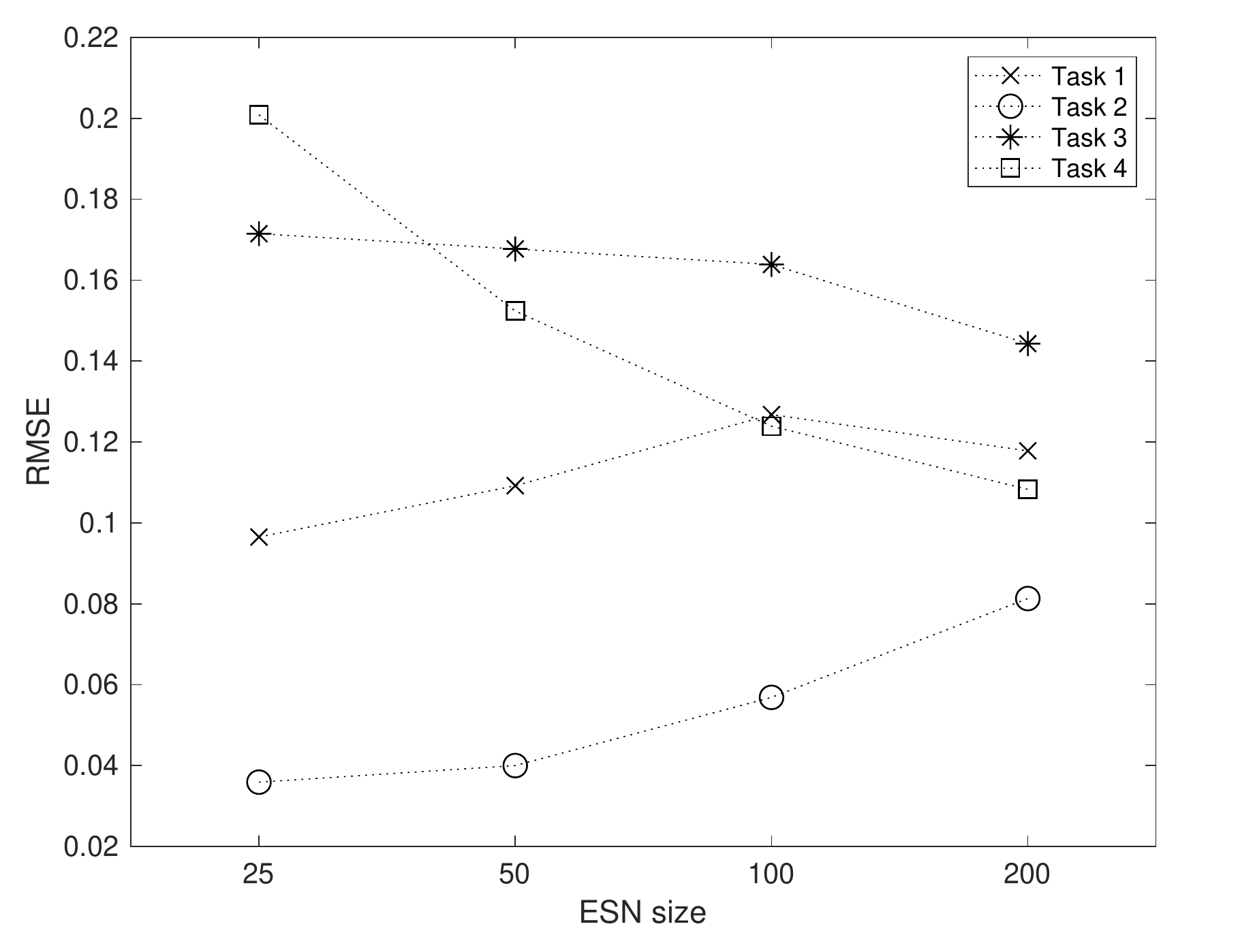}

\caption{Mean prediction error (RMSE) of FFNNs across all tasks and ESN sizes. Task 1: NARMA-10, Task 2: NARMA-30, Task 3: Santa Fe laser, and Task 4: non-linear channel equalisation. (The spread in values across the 10 FFNNs evaluated at each point is too small to see on this plot.) \label{fig: ffnn predict}}
\end{figure*}
%%%%%%%%%%%%%%%%%%%%%%%%%%% Figure %%%%%%%%%%%%%%%%%%%%%%%%%%%

To create the prediction model, four common reservoir computing benchmark tasks are selected: the non-linear autoregressive moving average (NARMA) task with a 10th and a 30th order time-lag; the Santa Fe laser time-series prediction task; the non-linear channel equalisation (NCE) task. Each task requires a different set of reservoir properties. Full details of the tasks are provided in Appendix~\ref{app: benchmarks}.

The modelling process of matching behaviours to task performances is framed as a regression problem. The model is created using standard feed-forward neural networks (FFNN) and trained using a sample of the behaviours discovered in the exploration process, and their evaluated task performances. The inputs of the FFNN are: MC (continuous-valued), KR and GR (discrete values). The output of the FFNN is the predicted task performance (continuous-valued) of each behaviour, measured as the normalised mean squared error (NMSE). 

The prediction error of the FFNN is measured on the test sample, as the root mean squared error (RMSE) between the predicted NMSE and the behaviour's actual evaluated NMSE for a given task.
That is, the prediction error $PE$ is
\begin{equation}
    PE = {\left(\frac{1}{N}\sum_{i \in test} \left(ptp_i - atp_i\right)^2\right)}^{1/2}
\end{equation}
where $N$ is the size of the test set, $ptp$ is the predicted task performance NMSE, and $atp$ is the actual task performance NMSE. 

In the experiment, multiple FFNNs of the same size are trained per task, and per substrate network size (see Appendix~\ref{app: accuracy BS} for experimental details). If the behaviour space  provides a reliable representation, the mean prediction error of the trained FFNNs should be low, since reliability implies a strong relationship is present, that not too difficult to model, and that is similar when network size changes. 

Some difference in prediction error is present between models trained with different network sizes. This is due to different behavioural ranges, resulting in an increase or decrease in complexity of the modelled relationships. For example, reservoirs in the behaviour space around $KR=GR=MC\leq 25$ tend to have similar poor performances for the NARMA-30 task because they do not meet a minimum requirement ($MC\geq 30$). This means the task is easier to model for small networks, as performance tends to be equally poor for all behaviours. Similarly, when larger ESNs are used to model the relationship, prediction error will likely increase as the distribution of errors increases and the complexity of the model increases. Patterns such as this are task-dependent, adding another level of complexity to the modelling process. For some tasks, to reliably model the relationship  requires a greater variety of behaviours than smaller ESNs can provide. Therefore, FFNNs trained on the behaviour space of 200 node network perform better than ones provided by the smaller networks, despite the apparent increase in complexity. 

The mean prediction errors of the FFNNs, for each task and substrate size, are shown in Fig.\ref{fig: ffnn predict}. Overall, the prediction errors are low, with typical values of $<0.16$. Depending on the task, errors increase or decrease as substrate network size increases. The prediction error for task 3 (Santa Fe laser) and task 4 (non-linear channel equalisation) decreases with substrate size, suggesting the model improves when trained using a larger variety of behaviours. However, these two tasks are particularly challenging to model (with a typical RMSE $>0.1$) because of outliers in the training data coming from poor (high task error) and very good (low task error) reservoirs, typically with an NMSE $\ll 0.1$. 

For the NARMA tasks, task 1 (NARMA-10) and 2 (NARMA-30), the prediction error increases as the network size increases. Prediction accuracy of the model therefore tends to degrade when trained with larger behaviour spaces, in contrast with tasks 3 and 4. 
However, this increase in error happens as variation in task performances increases; mirroring the same modelling problem for tasks 3 and 4. The lowest task errors for the NARMA-10 drop from an NMSE $\approx0.13$ to an NMSE $\approx0.01$ as size increases. The same also occurs for the NARMA-30 task, with the lowest errors decreasing from an NMSE $\approx0.48$ to an NMSE $\approx0.14$.

From these results, a strong correlation emerges between the variance in task performance (NMSE) of each behaviour space and the prediction error (RMSE). This suggests refocusing the learning process: instead of trying to reliably model \textit{all} behaviours, including the poor reservoirs, try to reliably model and predict only the best performing behaviours.  
The additional experiments in Appendix~\ref{app: accuracy BS} show the effect of this refocussing. The RMSE is significantly reduced when modelling behaviou\-rs below a task performance error threshold, rather than all behaviours. The results show the behaviour representation and model is most reliable when representing only the better task performing behaviours.

Overall, the results of this evaluation step suggest the behaviour space provides a sufficiently reliable representation of the substrate's computational capabilities. However, given that the provided behaviour measures are known not to capture all the interesting dynamics of the system, there is room to improve the behaviour representation and the modelling process.

%-------------------------------------------------
\section{Phase Two: Test Substrates}\label{sec:phase two}

Phase two of the framework, $P2$, evaluates the test substrate(s) against the phase one reference substrate(s). 
The behaviour space of the test substrate(s) is explored in level $P2.2$;
the quality is determined relative to the reference in level $P2.3$;
here the framework's substrate-independence is evaluated in level $P2.4$.

To demonstrate and evaluate the framework, two test substrates are characterised here: a simulated delay-based reservoir, and a physical carbon-nanotube based system. Each chosen substrate poses a unique challenge for the framework. These include differences in: implementation (simulated or physical), structure (spatial or temporal), and levels of noise in each system.

\subsection{Delay-based Reservoir (DR)}

The first substrate to be characterised is based on the Delay-Line Reservoir (DR) system~\cite{appeltant2011information,larger2012photonic,ortin2015unified}, using a single non-linear node and a delay line. This particular reservoir system mimics the structure of a recurrent network of coupled processing nodes in the time domain rather than spatially. By applying time multiplexing and nonlinear mixing to the input signal, a virtual network of processing nodes is created. To date, DR's have produced excellent performances across different reservoir computing benchmarks~\cite{appeltant2011information,duport2012all,paquot2012optoelectronic}.

Delay-feedback dynamical systems possess high-dim\-ensional state spaces and tunable memory making them ideal candidates for reservoir computing. The dynamics of these systems are typically modelled using delay differential equations of the type:

\begin{equation}
    \frac{d}{dt}x(t) = -x(t) + f(x(t-\tau),J(t)) \in \mathbb{R}
\end{equation}
where $t$ is time, $\tau$ is the delay time, $f$ is the non-linear function, and $J(t)$ is the weighted and time multiplexed input signal $u(t)$.

The DR technique is popular for optical and optoelectronic dynamical systems as it enables the exploitation of properties unique to these systems. It also provides a simple structure to overcome technical hardware challenges. These include: exploiting high-bandwidth and ultra high-speeds, and removing the demanding requirement of large complex physical networks. The technique however is not limited to these systems. It also offers a novel approach to implement networks efficiently on other hardware platforms. This is particularly useful when few inputs and outputs are available, creating the required state and network complexity in the time-domain to solve tasks. Examples include electronic circuits~\cite{appeltant2011information,soriano2015delay}, Boolean nodes on a field-progr\-ammable gate array (FPGA)~\cite{haynes2015reservoir}, a non-linear mechanical oscillator~\cite{dion2018reservoir}, and spin-torque nano-osc\-illators~\cite{torrejon2017neuromorphic}. 
However, the DR technique also has potential shortcomings including: a serialised input, pre-processing required on the input, and limits determined by the length of the delay line. To overcome some of these shortcomings, more complex architectures of multiple time-delay nodes have been proposed, leading to improved performances compared to single node architectures~\cite{ortin2017reservoir}.   

The DR system characterised here consists of a simulated Mackey-Glass oscillator and a delay line, inspired by~\cite{appeltant2011information}. This same system was also realised physically using an analogue electronic circuit in~\cite{appeltant2011information}. Details on the implementation of the Mackey-Glass system and the time-multiplexing procedure are provided in Appendix~\ref{app: DR appendix}.

% In the past, DR systems have been characterised with numerous property measures, including KR, GR, linear and non-linear memory capacity measures~\cite{paquot2012optoelectronic,duport2012all,appeltant2012reservoir}.

\subsection{Physical Carbon Nanotube-based Reservoir (CNT)}

The second substrate to be characterised is a physical material deposited on a micro-electrode array. The substrate is electrical stimulated and observed using voltage signals and configured through the selection of input and output locations on the array. The material is a mixed carbon nano\-tube--polymer composite, forming random networks of semi-conducting nanotubes suspended in a insulating polymer. 
The material has been applied to, and performs well on, several computational problems including function approximation, the travelling salesman problem, and robot control~\cite{clegg2014travelling,mohid2016evolution}. However, the material has so far produced only modest performances on challenging reservoir computing benchmark tasks~\cite{dale2016evolving}. As a reservoir, the material has been shown to perform well on simple tasks, but struggles to exhibit strong non-linearity and sufficient memory for more complex tasks~\cite{dale2016ICES,dale2017IJCNN}.

In previous work~\cite{dale2016evolving,dale2016ICES,dale2017IJCNN}, a small level of characterisation was carried out on different carbon nanotube-based (CNT) reservoirs, showing even the best fabricated material (a 1\% concentration of carbon nano\-tubes w.r.t.\ weight mixed with poly-butyl-methacrylate) typically exhibits low memory capacity, despite different biasing and stimulation methods for configuration. 

The right concentration and arrangement of carbon nanotubes, and method for stimulating and observing the material, is still an open question. So far, the methods and materials used have led to overall modest performances on benchmark tasks such as NARMA-10~\cite{dale2016evolving} and the Santa Fe laser time-series prediction task~\cite{dale2016ICES}, but encouraging when the number of inputs and outputs are taken into account. 

Characterising a black-box material like the carbon nanotube composite is challenging because of its disordered structure and stochastic fabrication process, making it impractical (or even impossible for the general case) to model its exact internal workings. Originally, the CNT-polymer composite was proposed as a sandpit material to discover whether computer-controll\-ed evolution could exploit a rich partially-constrained sour\-ce of physical complexity to solve computational problems~\cite{broersma2012nascence}. Because of its physicality, with somewhat unknown computing properties, it provides a challenging substrate for the CHARC framework to characterise. Further details about the carbon nanotube-based substrate and its parameters are provided in Appendix ~\ref{app: cnt appendix}.

\subsection{Quality of Test Substrates} \label{sec: quality of substrates}

\begin{figure}[tp]
\centering
\includegraphics[width=0.475\textwidth,trim=1.5cm 7cm 0.2cm 7cm,clip]{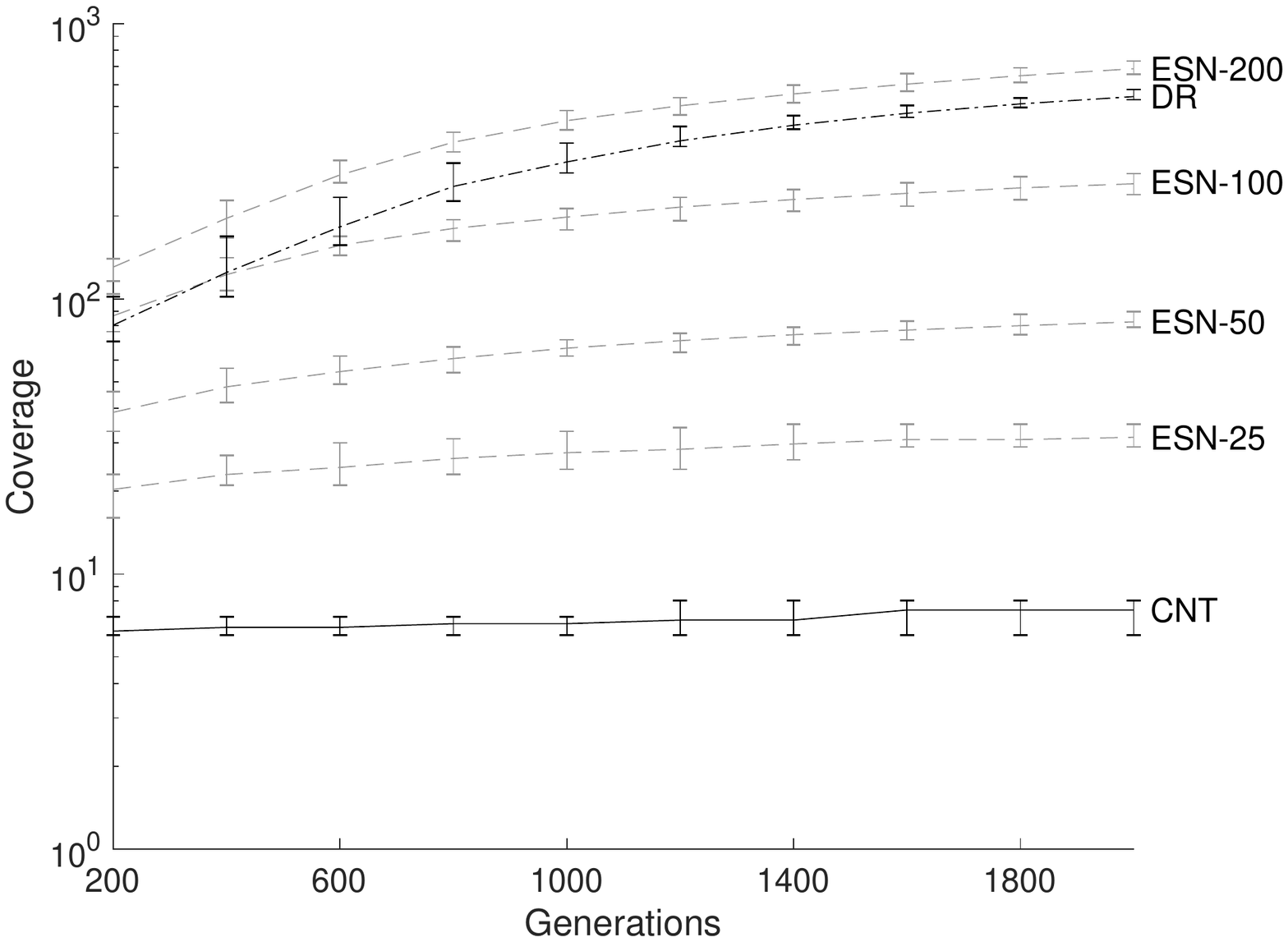}

\caption{Voxel measure of coverage as number of generations increase. Test substrates are shown as solid black lines, reference substrates are dashed grey lines. Error bars display the min-max values for all search runs. 
Note the logarithmic coverage scale.
\label{fig: phase two voxel results}}
\end{figure}

A visualisation of exploration level $P2.2$ and the results of the evaluation level $P2.3$ for each substrate are presented here. Similar to phase one, the quality of each substrate is calculated as the total number of voxels occupied after 2000 search generations. Fig.~\ref{fig: phase two voxel results} shows the total number of occupied voxels after every 200 generations, with error bars displaying the min-max values for different evolutionary runs. 

%%%%%%%%%%%%%%%%%%%%%%%%%%% Figure %%%%%%%%%%%%%%%%%%%%%%%%%%%
%%%%%%%%%%%%%%%%%%%%%%%%%%% Figure %%%%%%%%%%%%%%%%%%%%%%%%%%%

%After evaluating the initial population, 
The differences in behavioural freedom between the DR, CNT and ESN substrates are significant. Using the voxel measure, we can determine which of the reference substrates are close equivalents in terms of quality to the test substrates. At the beginning of the search process, the DR appears similar in quality to an ESN with 100 nodes, while the CNT has a considerably smaller quality than the ESN of 25 nodes. 
As the number of search generations increases, the DR's coverage increases rapidly, reaching a final value close to an ESN with 200 nodes, yet the CNT struggles to increase its coverage. The rate at which behaviours are discovered for the DR and CNT are very telling, suggesting it is much harder to discover new behaviours for the CNT than the DR. This increased difficulty could imply the bounds of the substrates behaviour space have almost been met: as the discovery rate of new novel behaviours decreases, either the search is stuck exploiting a niche area, or it has reached the boundaries of the whole search space.

% comparison to ESNs
%%%%%%%%%%%%%%%%%%%%%%%%%%% Figure %%%%%%%%%%%%%%%%%%%%%%%%%%%
\begin{figure*}[tp]
\centering
\subfloat[200 node ESN (light grey) with DR (dark grey)]{\includegraphics[width=0.85\textwidth,trim=2.9cm 6.2cm 3cm 7cm,clip]{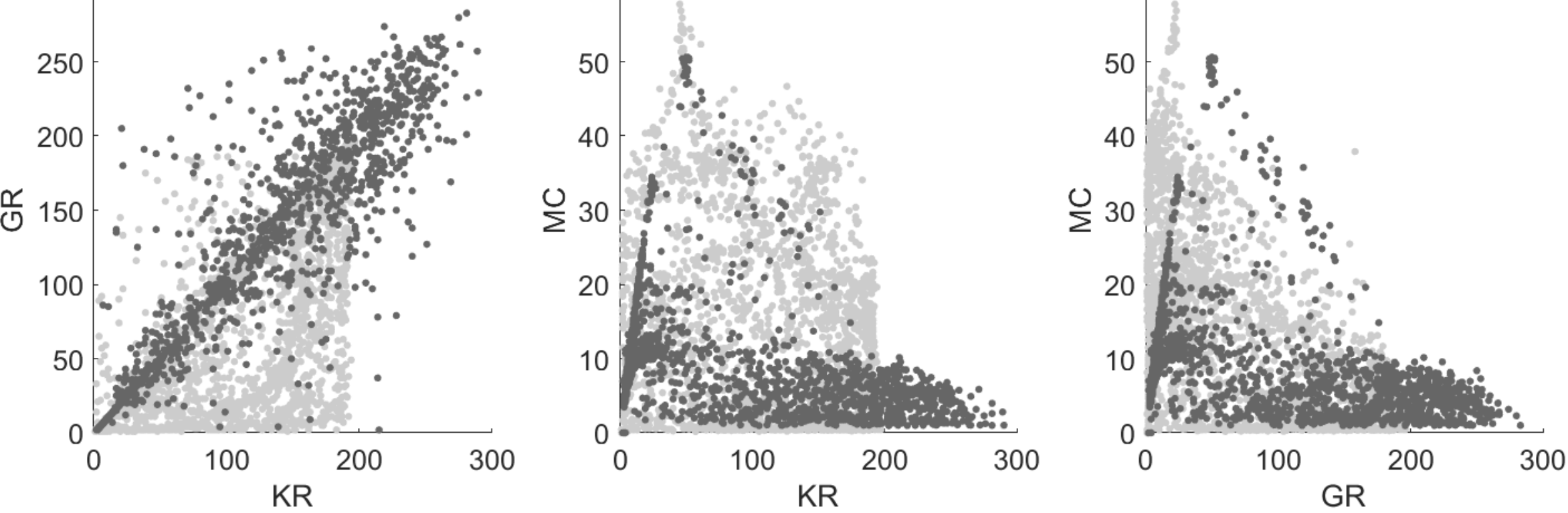}\label{fig: DR substrate}}

\subfloat[25 node ESN (light grey) with CNT (dark grey)]{\includegraphics[width=0.85\textwidth,trim=2.9cm 6.2cm 3cm 7cm,clip]{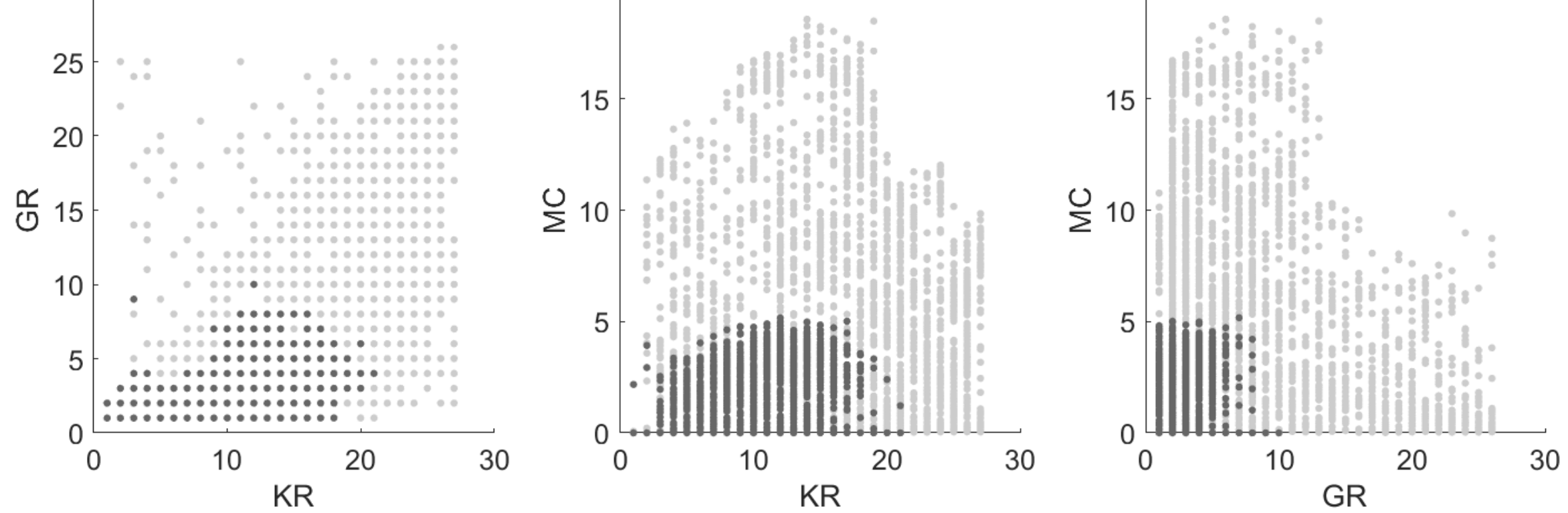}\label{fig: cnt substrate}}

\caption{Behaviours discovered when exploring the ESN, CNT \& DR substrates. To visually compare substrates, each test substrate is plotted over the reference substrate with the most similar quality. 
\label{fig: bs of substrates}}
\end{figure*}
%%%%%%%%%%%%%%%%%%%%%%%%%%% Figure %%%%%%%%%%%%%%%%%%%%%%%%%%%

A visual inspection of the covered behaviour spaces provides a more detailed understanding of the final quality values. The discovered behaviours for both substrates are shown in Fig.~\ref{fig: bs of substrates}. In each subplot, the behaviours for each test substrate (DR in Fig.~\ref{fig: DR substrate} and CNT in Fig.~\ref{fig: cnt substrate}) are presented in the foreground and reference substrates with the most similar quality (200 node ESN in Fig.~\ref{fig: DR substrate} and 25 node ESN in Fig.~\ref{fig: cnt substrate}) are placed in the background. 

In Fig.~\ref{fig: bs of substrates}a, the DR behaviours extend into regions that the 200 node ESN cannot reach, and as a consequence, only sparsely occupies regions occupied by the ESN. Given more search generations, these sparse regions would likely be filled, as similar behaviours are already discovered. 

The DR struggles to exceed the memory capacity of the 200 node ESNs, or exhibit a KR or GR beyond 300, despite having 400 virtual nodes. This could indicate that increasing the number of virtual nodes does not necessarily lead to greater memory or dynamical variation, a feature more typical of ESNs (see Fig.~\ref{fig: ESN BS overlay}). However, the virtual network size is not an isolated parameter; the time-scale and non-linearity of the single node, and the separation between virtual nodes, all play an important role in reservoir dynamics. 

In Fig.~\ref{fig: bs of substrates}b, the CNT exploration process struggles to find behaviours with $MC>5$, reaching what appears to be a memory capacity limit. The highest discovered KR and GR values are also small, tending to be lower than (almost half) their possible maximum values, i.e., the total number of electrodes used as outputs. This suggests the substrate struggles to exhibit enough (stable) internal activity to create a strong non-linear projection, and to effectively store recent input and state information; agreeing with previous results~\cite{dale2016evolving,dale2016ICES,dale2017IJCNN}. The results here also highlight why only a limited range of tasks are suitable for the substrate, and why small ESNs tend to be good models of the substrate.

These results show the CNT substrate in its current form features a limited set of behaviour, explaining its usefulness to only a small class of problems. 
The DR system features greater dynamical freedom, implying it can perform well across a larger set of problems. The coverage of this particular Mackey-Glass system is similar to large ESNs, explaining why they can closely match the performance of ESNs across the same class of problems~\cite{appeltant2011information,appeltant2014constructing}.

\subsection{Prediction transfer} \label{sec: pred across subs}

%%%%%%%%%%%%%%%%%%%%%%%%%%% Figure %%%%%%%%%%%%%%%%%%%%%%%%%%%
\begin{figure*}[tp]
\centering
\includegraphics[width=0.7\textwidth,trim=1.25cm 7.5cm 1.5cm 7cm,clip]{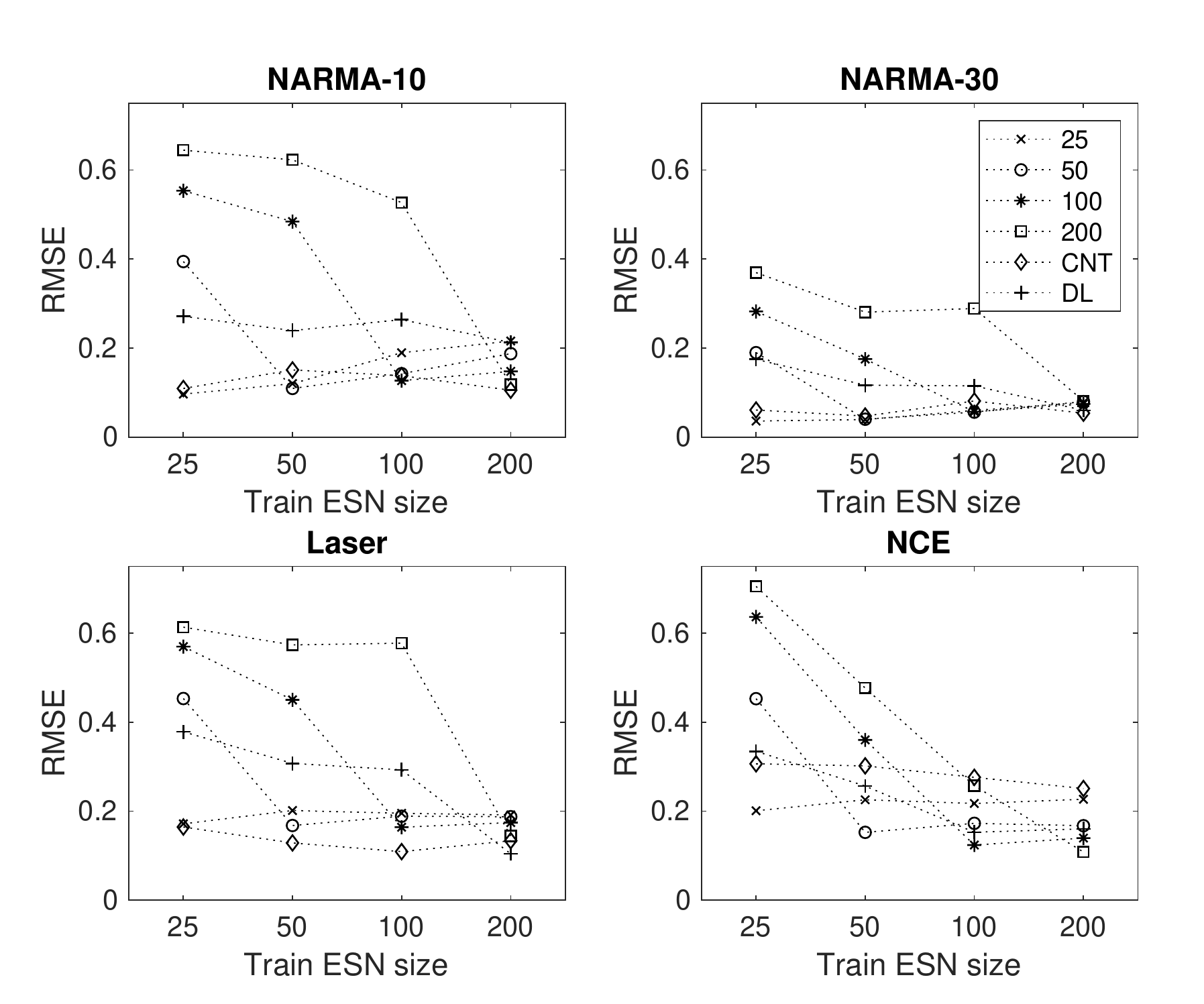}
\caption{Prediction error (RMSE) of the learned models (FFNNs) from $P1.3$
when set to predict the task performance of other substrates. The modelled reference substrate (i.e., ESN size) used for the prediction is given on the x-axis and the test substrate is given in the legend. \label{fig: ffnn predict material}}
\end{figure*}
%%%%%%%%%%%%%%%%%%%%%%%%%%% Figure %%%%%%%%%%%%%%%%%%%%%%%%%%%

The final level here, $P2.4$, evaluates the substrate-indep\-endence of the framework. To do this, we evaluate the transferability of the learnt relationships (FFNNs) from level $P1.4$ by measuring their prediction accuracy when tasked to predict a different substrate.
We evaluate how well the trained models (FFNNs) of the reference substrates predict the performance of the other reference substrates, i.e., predict the task performance of different ESN sizes.

Fig.~\ref{fig: ffnn predict material} shows the mean prediction error (RMSE) of all FFNNs for every predicted substrate. Each dashed line represents the predicted substrate. The $x$-axis represents the FFNNs trained on different reference network sizes; four sizes are shown for each task, being FFNNs trained using the ESN sizes 25, 50, 100 and 200 nodes. The $y$-axis is the prediction error (RMSE) of each model for each substrate. 

The results show that the models trained with smaller network sizes tend to poorly predict the task performance of larger networks across all tasks. This intuitively makes sense; the smaller network models are trained without any data examples beyond their own behavioural limits, and thus cannot make an accurate prediction for larger networks.

The models trained with larger network sizes tend to predict the smaller networks fairly well. The best predictions occur when the model is trained and tested using the same network size. Considering the variation in task performance as size increases, and fewer training examples within specific areas occupied by smaller network sizes, prediction appears to be reasonably robust when using the largest explored reference substrate. 

The model of the largest network (200 node) tends to better predict the DR, on average resulting in the lowest prediction errors. For the CNT, models of all network sizes result in low prediction errors for most tasks, except the non-linear channel equalisation task. Prediction error for this task, however, continues to improve as network size increases. Given these results, we argue that a reference substrate with high quality will tend to provide a good prediction of lower quality substrates.

\begin{figure*}[tp]
\centering
\includegraphics[width=0.7\textwidth,trim=1.25cm 7.5cm 1.5cm 8cm,clip]{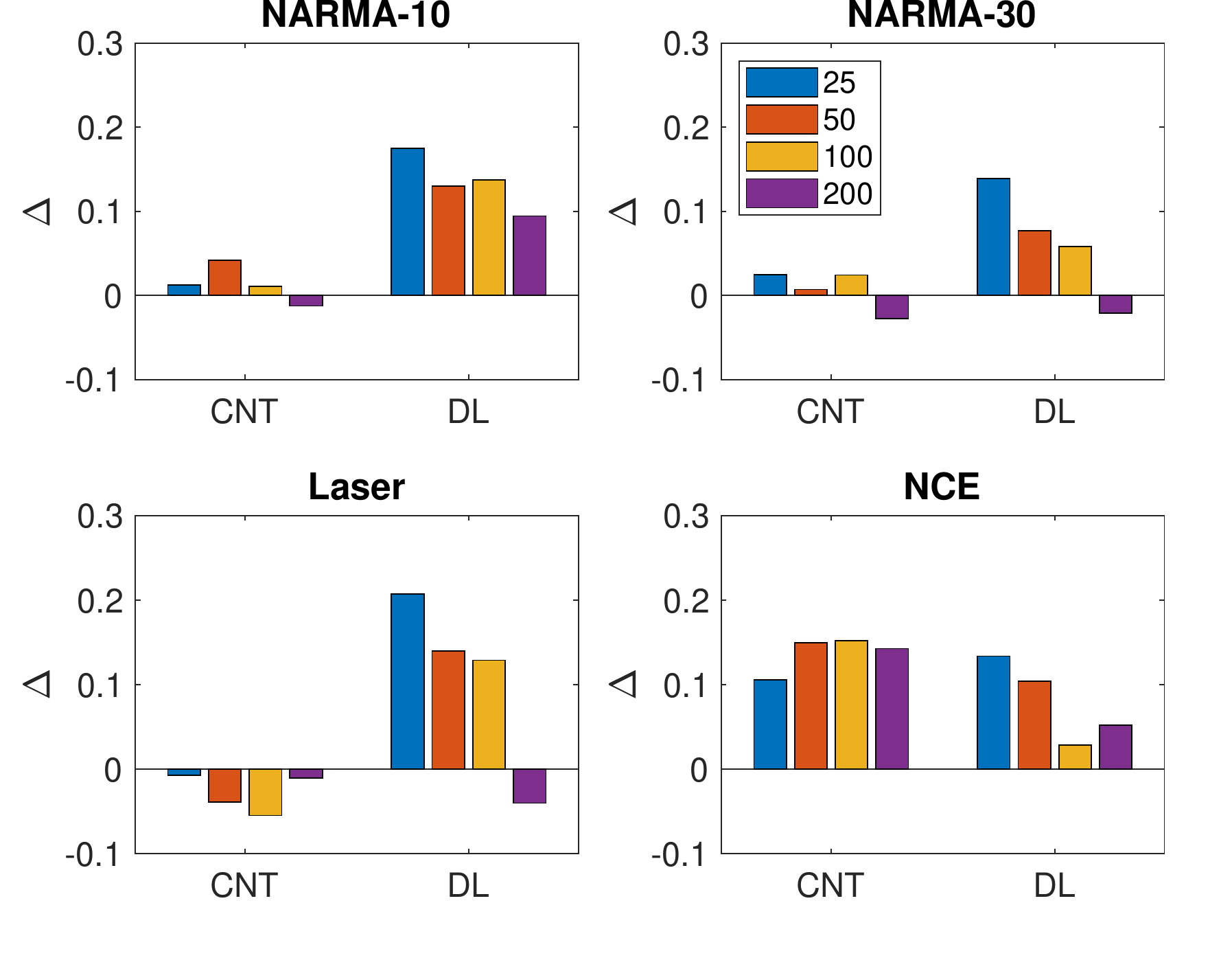}
\caption{Difference ($\Delta$) between best (self-)prediction and test prediction for CNT and DR substrates.\label{fig: delta prediciton plot}}
\end{figure*}

Fig.~\ref{fig: delta prediciton plot} summarises the results of the substrate-indepe\-ndence experiment. 
It plots the difference ($\Delta$) between the best prediction error and the test substrates prediction error.
When the overall prediction error is low and the difference ($\Delta$) is close to zero, the relationship between behaviour and task performance is strong, and thus the abstract behaviour space reliably represents underlying computational properties, independent of the substrate's implementation.

%%%%%%%%%%%%%%%%%%%%%%%%%%% Figure %%%%%%%%%%%%%%%%%%%%%%%%%%%
%%%%%%%%%%%%%%%%%%%%%%%%%%% Figure %%%%%%%%%%%%%%%%%%%%%%%%%%%

Fig.~\ref{fig: delta prediciton plot} plots $\Delta$ for the two test substrates on all four benchmark tasks. Each bar signifies the difference $\Delta$ between the best prediction error (from the model trained and tested with the same network size) and the trained model used to predict the test substrate. The results show on average the CNT tends to provide the smallest $\Delta$ with models of smaller networks. For the DR, the model of the largest network tends to provide $\Delta$'s closest to zero.

Overall, the low and similar prediction errors across substrates indicates that the CHARC framework has a good level of substrate independence. The results also highlight the non-trivial nature of modelling the task--property relationship, with some tasks being more difficult to model and predict than others. Although not the original purpose of this level, this demonstrates that one could roughly predict the task performances of newly characterised substrates, or potentially even test new tasks using a trained model without having to evaluate the test substrate directly. This feature of the framework is potentially beneficial to hardware systems where training can be time and resource intensive.

\section{Conclusion}

A fundamental question in reservoir computing is: for a given task, what characteristics of a dynamical system or substrate are crucial for information processing? The CHARC framework tackles this question by focusing on the characteristic behaviour of the substrate rather than its performance on a specific task. In the process, two non-trivial problems were attempted; (i) how to characterise the quality of a substrate for reservoir computing; and (ii) how do computational properties relate to performance. 

To utilise the framework, two phases must be completed. In the first phase, the basic levels (\textit{definition}, \textit{exploration} and \textit{evaluation}) are applied to a reference substrate, providing context for future quality characterisations for other substrates. In the second phase the test substrate is explored, characterised and compared.

The presented framework is flexible, allowing new computational measures, techniques and additional high-level functions to be added. In this work, we have proposed and demonstrated just one possible high-level function that could model the challenging relationships between tasks and computational properties. This is used to predict the task performance of the substrate given its task-independent behaviour.

Using the framework, we have shown that exploration through open-ended evolution can be a powerful tool for outlining the limitations and capability of a substrate. This explains why a carbon nanotube-based composite can solve some simple computational tasks but often struggles to compete with more complex reservoir substrates. It is also shown why delay-based reservoirs compare so favourably to echo state networks due to similar behavioural quality. 

The characterisation process of CHARC has many potential future applications, for example: assessing the effect structure, topology and complexity has on dynamical freedom; using quality to guide, understand and explore substrate design; and, eventually, the design of suitable computational models. Ultimately, this can open the door for the co-design of both computational model and substrate to build better, more efficient unconventional computers.

\section*{Acknowledgements}
This work was part-funded by a Defence Science and Technology Laboratory (DSTL) PhD studentship, and part-funded by the SpInsired project, EPSRC grant EP/R032823/1. 

\bibliographystyle{abbrv}
\bibliography{reservoir}

\newpage
\appendix
\section{Novelty Search} \label{app: novelty search}

In the presented framework, an open-ended evolutionary algorithm called novelty search (NS) \cite{lehman2008exploiting,lehman2010efficiently,lehman2011abandoning} is used. Novelty search is used to characterise the substrate's behaviour space, i.e. the dynamical freedom of the substrate, by sampling its range of dynamical behaviours. 
In contrast to objective-based techniques, a search guided by novelty has no explicit task-objective other than to maximise novelty. Novelty search directly rewards divergence from prior behaviours, instead of rewarding progress towards some objective goal. 

Novelty search explores the behaviour space by promoting configurations that exhibit novel behaviours. Novelty of any individual is computed with respect to its distance from others in the behaviour space. To track novel solutions, an \textit{archive} is created holding previously explored behaviours. Contrary to objective-based searches, novelty takes into account the set of all behaviours previously encountered, not only the current population. This enables the search to keep track of (and map) lineages and niches that have been previously explored. 

To promote further exploration, the archive is dynamically updated with respect to two parameters, $\rho_{min}$ and an update interval. The $\rho_{min}$ parameter defines a minimum threshold of novelty that has to be exceeded to enter the archive. The update interval is the frequency at which $\rho_{min}$ is updated. Initially $\rho_{min}$ should be low, and then raised or lowered if too many or too few individuals are added to the archive in an update interval. Typically in other implementations, a small random chance of any individual being added to the archive is also set. 

In the presented implementation, a small initial $\rho_{min}$ is selected relative to the behaviour space being explored and updated after a few hundred generations. $\rho_{min}$ is dynamically raised by 20\% if more than 10 individuals are added and $\rho_{min}$ is lowered by 5\% if no new individuals are added; these values were guided by the literature~\cite{lehman2008exploiting}.

To maximise novelty, a selection pressure rewards individuals occupying sparsely populated regions in the behaviour space. To measure local sparsity, the average distance between an individual and its $k$-nearest neighbours is used. A region that is densely populated results in a small value of the average distance, and in a sparse region, a larger value. The sparseness $\rho$ at point $x$ is given by:
\begin{align}\label{eq: }
	\rho(x) = \frac{1}{k}\sum_{i=1}^{k} dist(x, \xi_i)
\end{align}
where $\xi_i$ are the $k$ nearest neighbours of $x$.

The search process is guided by the archive contents and the current behaviours in the population, but the archive does not provide a complete picture of all the behaviours explored. Throughout the search, the population tends to meander around existing behaviours until a new novel solution exceeding the novelty threshold is discovered. To take advantage of this local search, all the explored behaviours are stored in a separate database $D$. This database stores all the information used to characterise the substrate's later quality and has no influence on the search, which uses only the archive.

\subsection{Novelty Search Implementation}\label{sec: ns implementation}
% GA - description
In the literature, Novelty Search is frequently combined with the Neural Evolution of Augmented Topologies (NEAT) \cite{lehman2011abandoning,stanley2002evolving} representation; this neuro-evolutionary method focuses on adapting network topology and complexifying a definable structure. For the CHARC framework, a more generic implementation is needed, featuring a minimalistic implementation not based on any specific structure or representation. For this reason, an adaptation of the steady-state Microbial Genetic Algorithm (MGA) \cite{harvey2009microbial} combined with novelty search is used. The MGA is a genetic algorithm reduced to its basics, featuring horizontal gene transfer (through bacterial conjugation) and asynchronous changes in population where individuals can survive long periods. 

\begin{figure}[t]
\centering
\small
\begin{minipage}{.8\linewidth}
\begin{algorithm}[H]
    \caption{Novelty search with microbial GA algorithm}
    \label{alg:noveltyGA}
    \begin{algorithmic}
        \State $pop \gets random$ \Comment{initial random population list length $P$}
        \State $A \gets pop$ \Comment{archive initialised}
        \State $D \gets pop$ \Comment{database initialised}
        \While{searching}
        	\State $i :\in 1..popSize$ \Comment{parent 1 from pop}
        	\State $j :\in \mbox{deme}~ i$ \Comment{parent 2 from deme}
            \If{$f(pop(i),A,pop) > f(pop(j),A,pop)$} 
            	\State $winner, loser \gets i,j$ \Comment{fitness is novelty}
            \Else
            	\State $winner, loser \gets j,i$
            \EndIf
            \State $child \gets infection(winner,loser)$
            \State $child \gets mutation(child)$
            \State $pop(loser) \gets child$
            \If {$child \mbox{ is sufficiently novel}$}
            	\State $\mbox{add $child$ to $A$}$ 
            \EndIf
            \State $\mbox{add $child$ to $D$}$
            \If {generation == $n \times update_{gen}$} 
            	\State $\mbox{update novelty threshold }\rho_{min}$ 
            \EndIf
            %gen = gen + 1
        \EndWhile
    \end{algorithmic}
\end{algorithm}
\end{minipage}
\end{figure}

To apply the MGA to the problem a number of adaptations are required. Caching fitness values in the standard steady-state fashion is not possible, as fitness is relative to other solutions found and stored in the growing archive. In this implementation, no individual fitness's are stored across generations, however the same steady-state population dynamics are kept, i.e. individuals are not culled, and may persist across many generations. 

%%%%%%%%%%%%%%%%%%%%%%%%%%% Figure %%%%%%%%%%%%%%%%%%%%%%%%%%%
\begin{figure*}[tp]
\vspace{0.5cm}
\centering
\includegraphics[width=0.75\textwidth]{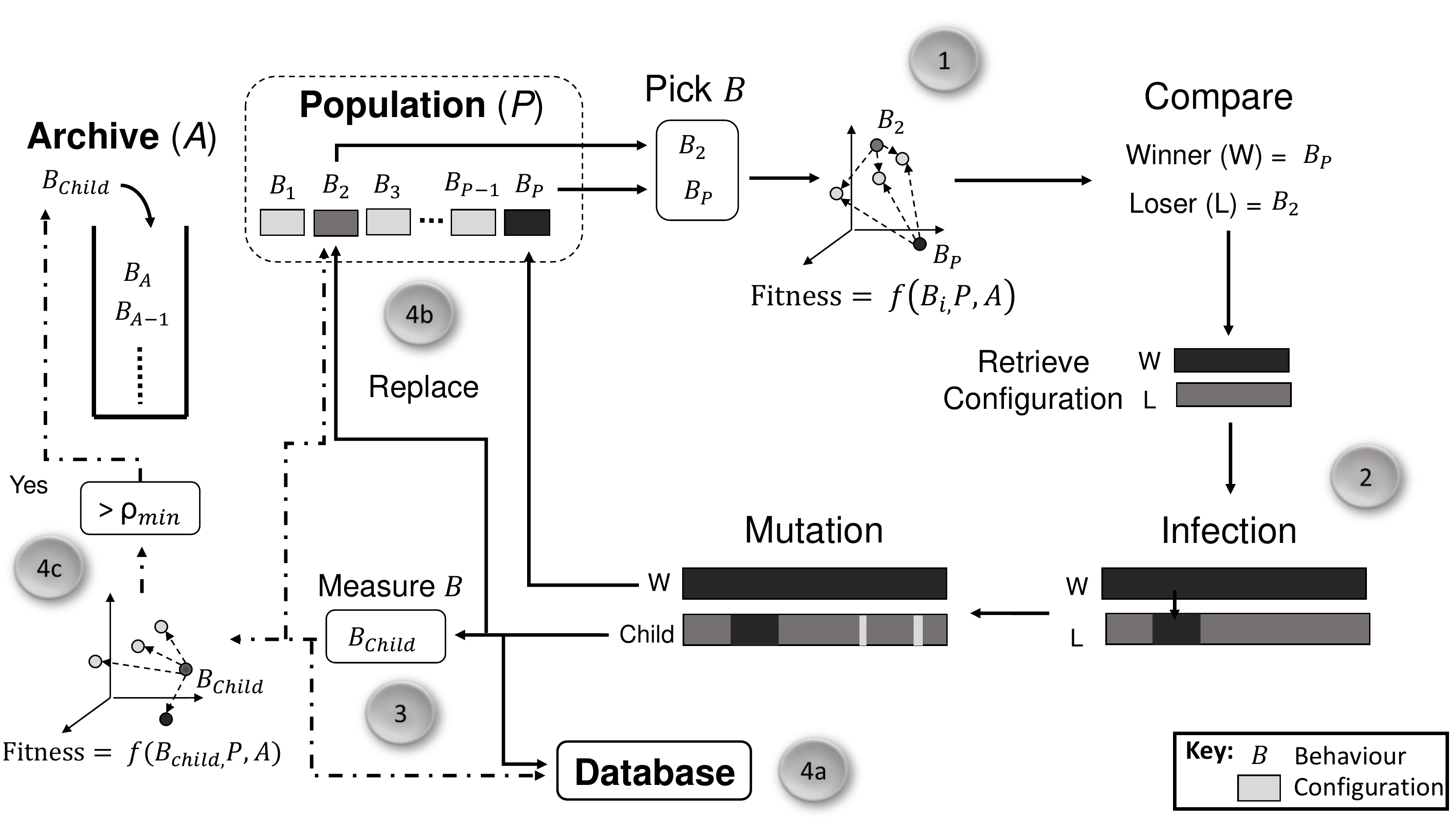} %plots/micro_GA_implementation
\caption{Adapted microbial GA with novelty search.
}
\label{fig: example microbial GA}
\end{figure*}
%%%%%%%%%%%%%%%%%%%%%%%%%%% Figure %%%%%%%%%%%%%%%%%%%%%%%%%%%

An overview of the evolutionary loop is given in Fig.~\ref{fig: example microbial GA}. The complete process is also outlined in pseudo-code in Algorithm~\ref{alg:noveltyGA}.

At the beginning of the search process, a random population is created. In the population, both the substrate configurations and the resulting behaviours $B$ are stored. This initial population is then added to the archive $A$ and database $D$. 

At step $1$, tournament selection with a tournament size of two is used. To ensure speciation, the first parent is picked at random and the second is chosen within some proximity to the other determined by the MGA parameter \textit{deme size}. In this step, the fitness values (novelty) of both behaviours are calculated relative to population $P$ and archive $A$. The individual with the larger distance, that is, occupying the less dense region of the behaviour space, is adjudged the winner. This elicits the selection pressure towards novel solutions. The microbial GA differs from other conventional GAs as the weaker (here, less novel) individual becomes ``infected'' by the stronger (more novel) one, replacing its original self in the population.

At step $2$, the configurations of both behaviours are retrieved and manipulated. This constitutes the infection and mutation phase. In the infection phase, the weaker parent undergoes horizontal gene transfer becoming a percentage of the winner and loser. The genetic information of the weaker parent does not disappear in this process, as some percentage defined by the recombination rate parameter remains intact. In the mutation phase, the weaker parent undergoes multiple point-mutations, becoming the new offspring. 

At step $3$, the configuration of the new offspring is untested, therefore the behaviour $B_{Child}$ of the individual needs to be updated. At steps $4a$ and $4b$, the offspring's behaviour and configuration are added to the database $D$ and it replaces the loser in the population $P$.

At the last step $4c$, the fitness/novelty of the offspring $B_{Child}$ is compared to both the current population $P$ and archive $A$. If the novelty of the offspring exceeds the novelty threshold $\rho_{min}$, the behaviour $B_{Child}$ (configuration is not needed) is added to the archive $A$. 

Overall, three fitness values are calculated at each generation. Two fitness evaluations occur in the selection phase and a third fitness evaluation is carried out on the offspring, in order to update the archive. The computational complexity of the fitness function is $O(nd+kn)$ using an exhaustive $k$-nearest neighbour search. As the dimension $d$ of the archive/behaviour space is small ($d=3$ property measures), the number of $k$-neighbours (here $k=15$) has the dominant effect. This value of $k$ is chosen experimentally; larger $k$-values improve accuracy but increase run time. As the archive size increases, run time increases proportional to archive size $n$. To reduce complexity, Lehman and Stanley~\cite{lehman2011abandoning} describe a method to bound the archive using a limited stack size. They find that removing the earliest explored behaviours, which may result in some limited backtracking, often results in minimal loss to exploration performance. 

The same NS parameters are applied to every substrate. These are: generations limited to 2,000; \textit{population size} = 200; \textit{deme} = 40; \textit{recombination rate} = 0.5; \textit{mutation rate} = 0.1; $\rho_{min}=3$; and $\rho_{min}$ \textit{update} = 200 generations. Five evolutionary runs are conducted for the CNT and delay-based reservoir, as the time to train increases significantly, and ten runs for the ESN substrates. 

\section{Benchmark Tasks for Prediction Phase} \label{app: benchmarks}

The NARMA task evaluates a reservoir's ability to model an \textit{n}-th order highly non-linear dynamical system where the system state depends on the driving input and state history. The task contains both non-linearity and long-term dependencies created by the \textit{n}-th order time-lag. An $n$-th ordered NARMA task predicts the output \(y(n+1)\) given by Eqn.~\eqref{eq:10thNarma} when supplied with \(u(n)\) from a uniform distribution of interval [0, 0.5].
For the 10-th order system parameters are: $\alpha =0.3, \beta=0.05, \delta =0.1$, and $\gamma =1$; for the 30-th order system: $\alpha =0.2, \beta=0.004, \delta =0.001$, and $\gamma =1$.

\ifisProcRoySocA
\begin{equation}\label{eq:10thNarma}
y(t+1) = \gamma\Biggr(\alpha y(t)+\beta y(t)\left(\sum_{i=0}^{n-1}y(t-i)\right) + 1.5u(t-9)u(t)+\delta\Biggr)
\end{equation}
\else
\begin{multline}\label{eq:10thNarma}
y(t+1) = \gamma\Biggr(\alpha y(t)+\beta y(t)\left(\sum_{i=0}^{n-1}y(t-i)\right) \\ 
+ 1.5u(t-9)u(t)+\delta\Biggr)
\end{multline}
\fi

The laser time-series prediction task predicts the next value of the Santa Fe time-series Competition Data (dataset A)\footnote{Dataset available at UCI Machine Learning Repository~\cite{LaserData}.}. The dataset holds original source data recorded from a Far-Infrared-Laser in a chaotic state. 

The Non-linear Channel Equalisation task introduc\-ed in \cite{jaeger2004harnessing} has benchmarked both simulated and physical reservoir systems~\cite{paquot2012optoelectronic}. The task reconstructs the original \textit{i.i.d} signal \(d(n)\) of a noisy non-linear wireless communication channel, given the output \(u(n)\) of the channel. To construct reservoir input \(u(n)\) (see Eqn.~\ref{eq: nonChanEQ pt2}) \(d(n)\) is randomly generated from ${-3, -1, +1, +3}$ and placed through Eqn.~\ref{eq: nonChanEQ pt1}:
%\begin{equation} 
\begin{align}\label{eq: nonChanEQ pt1}
q(n) = { }&0.08d(n+2)-0.12d(n+1) + d(n)  \notag\\
&{}+ 0.18d(n-1)-0.1d(n-2)\\
&{}+0.091d(n-3)-0.05d(n-4)\notag\\
&{}+0.04d(n-5)+ 0.03d(n-6) + 0.01d(n-7)\notag
\end{align}
%\end{equation}
\begin{equation} \label{eq: nonChanEQ pt2}
u(n) = q(n) + 0.036q(n)^{2} - 0.011q(n)^{3}
\end{equation}

Following~\cite{jaeger2004harnessing}, the input $u(n)$ signal is shifted +30 and the desired task output is $d(t-2)$.

\section{Substrate Parameters}\label{app:params}

\subsection{Echo State Networks}

In phase one, regardless of network size the same restrictions are placed on global parameter ranges and weights, applying the same weight initiation processes each time. For example, global parameters ranges include: an internal weight matrix ($W$) scaling between [0, 2], scaling of the input weight matrix ($W_{in}$) between [$-$1, 1], and the sparseness of $W$ [0, 1]. For both random and novelty search, at creation, a reservoir has each global parameter drawn from a uniform random distribution, as well as input weights and internal weights drawn uniformly from other ranges; $W_{in}$ between [$-$1, 1] and $W$ between [$-$0.5, 0.5].

\subsection{Carbon nanotube--polymer}  \label{app: cnt appendix}

The training and evaluation of the carbon-based substrate is conducted on a digital computer. Inputs and representative reservoir states are supplied as voltage signals. The adaptable parameters for evolution are the number of input-outputs, input signal gain (equivalent to input weights), a set of static configuration voltages (values and location), and location of any ground connections. Configuration voltages act as local or global biases, perturbing the substrate into a dynamical state that conditions the task input signal. 

% An advantage of physical substrate-based reservoirs is that computational speed of a trained reservoir is limited only by interface hardware and physical response time, which can potentially be all analogue. However, the training process can take considerably longer as multiple runs for statistical tests are often needed. With this substrate in particular, there are many unstable configurations. Therefore, extra evaluations are required to measure statistical stability in order to discard unstable signals from the training process. For this reason alone, the ability to predict performance across substrates and bypass the task training process would be decidedly beneficial. 

% \subsubsection{Experimental Parameters}
% The same NS parameters applied to the virtual ESN substrate are reused with all other substrates. These are: generations limited to 2,000; \textit{population size} = 200; \textit{deme} = 40; \textit{recombination rate} = 0.5; \textit{mutation rate} = 0.1; $\rho_{min}=3$; and $\rho_{min}$ \textit{update} = 200 generations. Five runs are conducted for the CNT and delay-based reservoir, as the time to train increases significantly. Ten runs are given for the ESN substrates. 

%%%%%%%%%%%%%%%%%%%%%%%%%%% Figure %%%%%%%%%%%%%%%%%%%%%%%%%%%
\begin{figure}[t]
\centering
\includegraphics[width=0.45\textwidth]{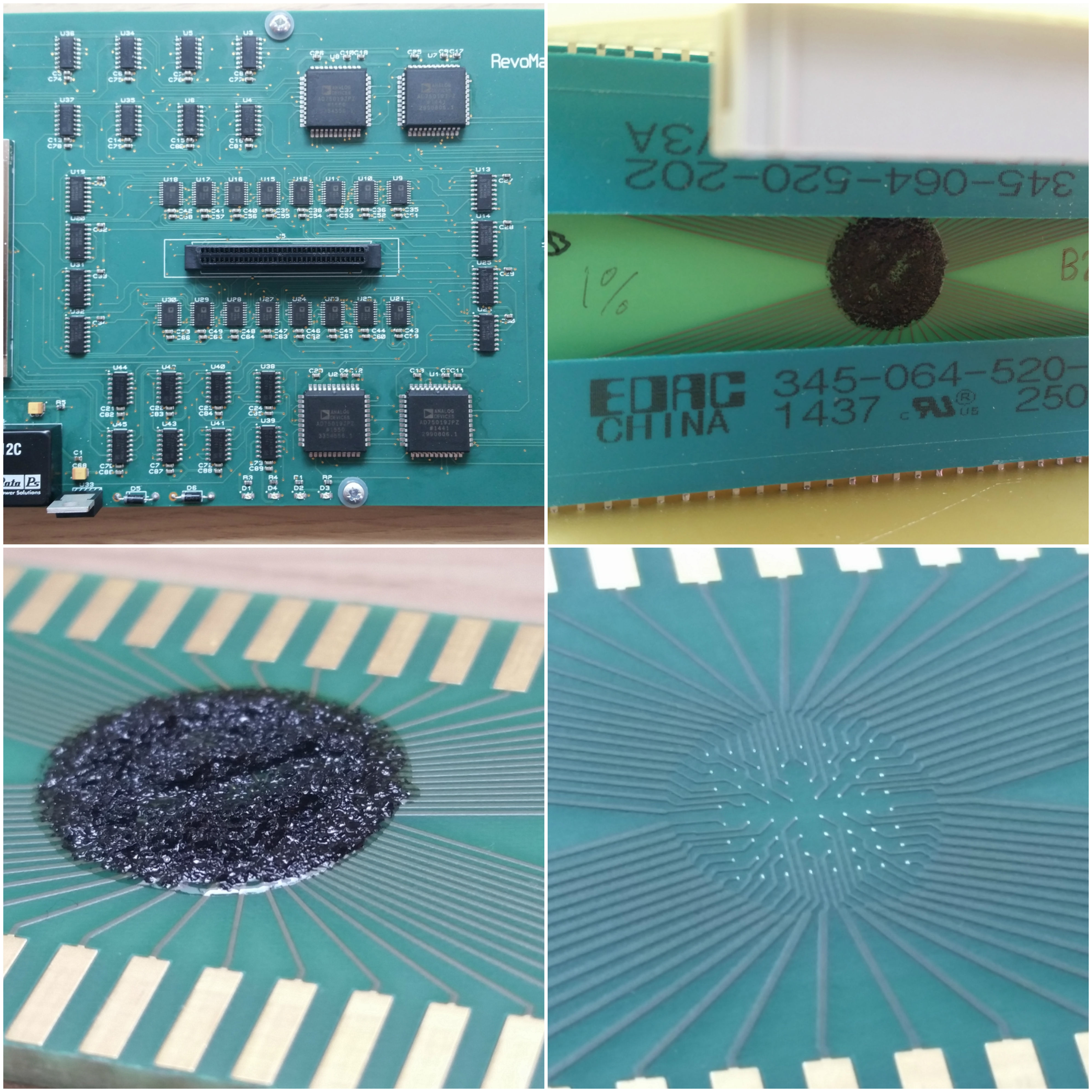}
\caption{Hardware Reservoir System. Micro-electrode housing, routing switch board and CNT--polymer deposited onto PCB electrode array.\label{fig: hardware system}}
\end{figure}
%%%%%%%%%%%%%%%%%%%%%%%%%%% Figure %%%%%%%%%%%%%%%%%%%%%%%%%%%

In this work, a 1\% carbon nano\-tube poly-butyl-methacryl\-ate (CNT/PBMA) mixture substrate is investigated. The substrate was mixed and drop cast onto a micro-electrode array using the same process in~\cite{dale2016evolving,dale2016ICES,dale2017IJCNN}. The electrode array comprises of 64 electrodes (contact sizes of 100$\mu$m and spacings of 600$\mu$m between contacts) deposited onto a FR-4 PCB using a chemical process that places Nickel and then a layer of Gold (see Fig.~\ref{fig: hardware system}).

Two National Instruments DAQ cards perform measurements and output analogue voltages; a PCI-6225 (16-Bit, 250 KS/s, with 80 analogue inputs), and PCI-6723 (13-Bit, 800KS/s, with 32 analogue outputs). Both cards communicate to a desktop PC through a session-based interface in MATLAB. The PCI-6723 supplies an additional 8 digital I/O lines to a custom routing board to program on-board switches and synchronise the cards. 

\subsection{Delay-based Reservoir} \label{app: DR appendix}

To generate $N$ virtual nodes and collapse them into a usable state observation, time-multiplexing is used. The input signal $u(t)$ is sampled and held for the period $\tau$ (the length of the delay line) and mixed with a random binary mask $M$, perturbing the node away from the relaxed steady state. For an interval defined by the node separation $\theta = \frac{\tau}{N}$, the mask is applied as a piecewise constant, forming the input to the non-linear node $J(t) = I(t)*M$. The state of the $i$-th virtual node is obtained after every $\tau$, as: $x_i (t) = x(\tau - (N-i)\theta)$.  

The model of the Mackey Glass dynamical system is described as:
\begin{align} \label{eq: DCR model}
    \dot{X}(t) = - X(t) + \frac{\eta \cdot [X(t-\tau) + \gamma \cdot J(t) ]}{1+[X(t-\tau)+\gamma \cdot J(t)]^{p}}
\end{align}
where $X$ represents the state, $\dot{X}$ its derivative with respect to time, and $\tau$ is the delay of the feedback loop. The parameters $\eta$ and $\gamma$ are the feedback strength and input scaling. The exponent $p$ controls the non-linearity of the node. The parameter $T$, typically omitted from Eq.~\ref{eq: DCR model}, represents the characteristic time-scale of the non-linear node. In order to couple the virtual nodes and create the network structure, $T\geq\theta$ is required. Together, all these parameters determine the dynamical regime the system operates within.

The parameters of the delay-based reservoir in this work are fixed at: $T=1$, $\theta=0.2$, $\tau=80$ and $N=400$, based on values given in~\cite{appeltant2011information}. During the exploration process, evolution can alter the mask, flipping between the binary values [-0.1, 0.1], and manipulate all of the Mackey Glass parameters: $0<\eta<1$, $0<\gamma<1$ and the exponent $0<p<20$.

\section{Property Measures}  \label{app: metrics}

\subsection{Kernel and Generalisation Rank}

The kernel measure is performed by computing the rank \(r\) of an \(n \times m\) matrix \(M\), outlined in \cite{busing2010connectivity}. To create the matrix \(M\), \(m\) distinct input streams \(u_{i},...,u_{m}\) are supplied to the reservoir, resulting in \(n\) reservoir states \(x_{u_{i}}\). Place the states \(x_{u_{i}}\) in each column of the matrix \(M\) and repeat \(m\) times. The rank \(r\) of \(M\) is computed using Singular Value Decomposition (SVD) and is equal to the number of non-zero diagonal entries in the unitary matrix. The maximum value of \(r\) is always equal to the smallest dimension of $M$. To calculate the effective rank, and better capture the information content, remove small singular values using some high threshold value. To produce an accurate measure of kernel rank \(m\) should be sufficiently large, as accuracy will tend to increase with \(m\) until it eventually converges.

The generalisation rank is a measure of the reservoir's capability to generalise given similar input streams. It is calculated using the same rank measure as kernel quality, however each input stream \(u_{i+1},...,u_{m}\) is a noisy version of the original \(u_{i}\). A low generalisation rank symbolises a robust ability to map similar inputs to similar reservoir states.

\subsection{Memory Capacity}

A simple measure for the linear short-term memory capacity (MC) of a reservoir was first outlined in~\cite{jaeger2001short} to quantify the \textit{echo state} property. For the echo state property to hold, the dynamics of the input driven reservoir must asymptotically wash out any information resulting from initial conditions. This property therefore implies a fading memory exists, characterised by the short-term memory capacity.

To evaluate memory capacity of an \(N\) node reservoir, we measure how many delayed versions \(k\) of the input \(u\) the outputs \(y\) can recall, or recover with precision. Memory capacity \(MC\) is measured by how much variance of the delayed input \(u(t-k)\) is recovered at \(y_k(t)\), summed over all delays. 
\begin{align} \label{eq: MC}
MC = \sum_{k = 1}^{2N} MC_{k} = \sum_{k=1}^{2N} \frac{cov^{2}(u(t-k),y_k(t))}{\sigma^{2}(u(t))\sigma^{2}(y_k(t))}
\end{align}

A typical input consists of \(t\) samples randomly chosen from a uniform distribution between [0 1]. Jaeger \cite{jaeger2001short} demonstrates that echo state networks driven by an i.i.d.\ signal can possess only $MC \leq N$. 

A full understanding of a reservoir's memory capacity cannot be encapsulated through a linear measure alone, as a reservoir will possess some non-linear capacity. Other memory capacity measures proposed in the literature quantify the non-linear, quadratic and cross-memory capacities of reservoirs~\cite{dambre2012information}. 

\section{Effect of Voxel size}\label{app:voxel}

To evaluate quality, a simple voxel based measure is used. 

% Three examples of how voxel size affects quality are shown in Fig.~\ref{fig: voxel measures}. Here, the same reference networks are measured using voxel sizes: $V_{size} = 1^3$, $V_{size} = 5^3$ and $V_{size} = 10^3$. 

% With $V_{size} = 1^3$, the networks explored with Novelty Search appear to have higher quality, resulting in a drastic overestimation. To query this deceptive value, we plot the discovered behaviours for three networks; two randomly explored and one explored via Novelty Search. According to the voxel measure with $V_{size} = 1^3$, a network of 50 nodes appears to have greater dynamical freedom than the larger networks. Yet, this is clearly not the case when looking at the behaviour space. As voxel size increases, from $V_{size} = 1^3$ to $V_{size} = 10^3$, we see a more realistic approximation of quality when compared with the visual inspection.

\begin{figure*}[t]
\centering
\subfloat[$V_{size}=1\times1\times1$]{\includegraphics[width=0.3\textwidth,trim=1.5cm 7.5cm 0cm 7cm,clip]{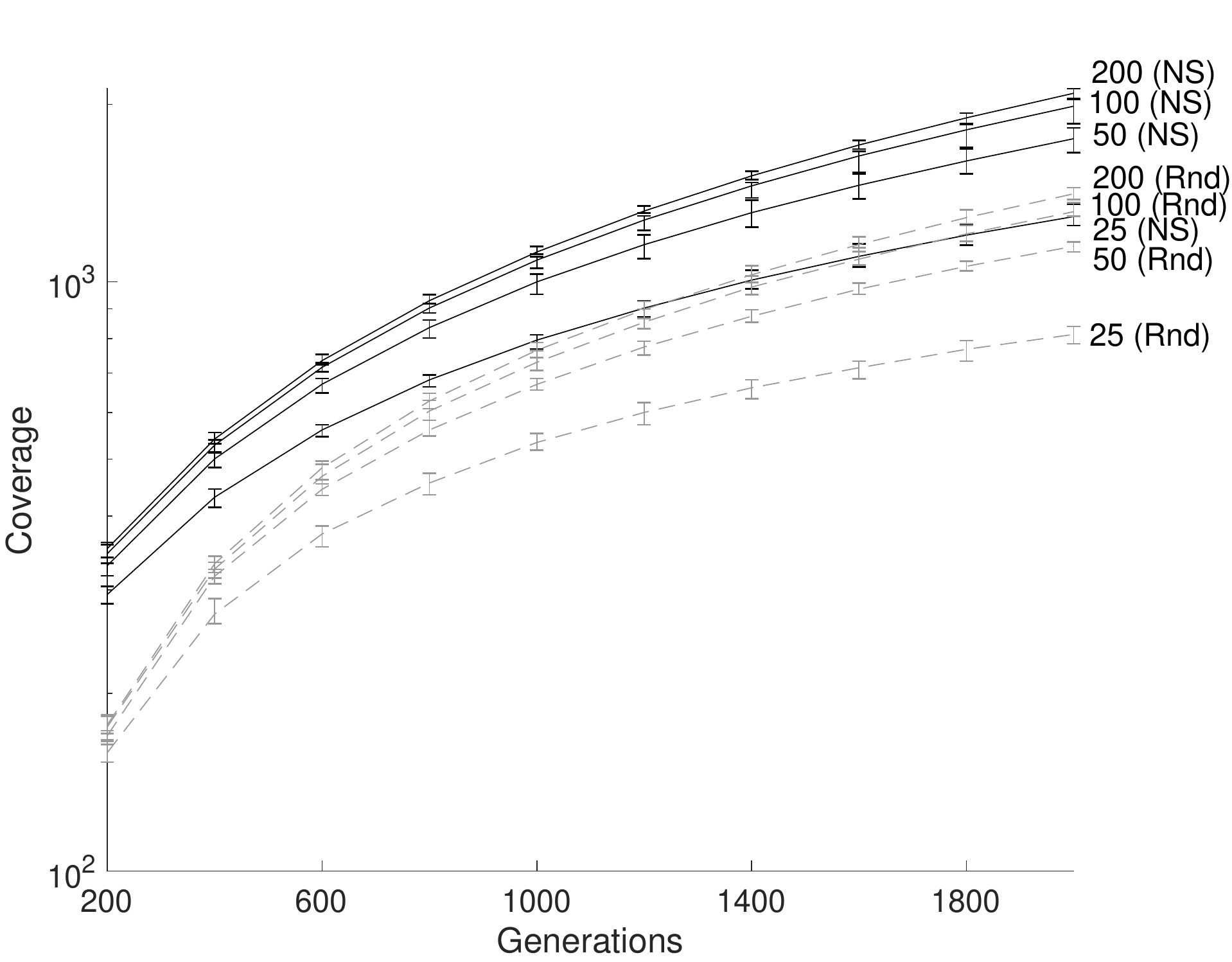}\label{fig: 1x1x1 voxel}} 
\subfloat[$V_{size}=5\times5\times5$]{\includegraphics[width=0.3\textwidth,trim=1.5cm 7.5cm 0cm 7cm,clip]{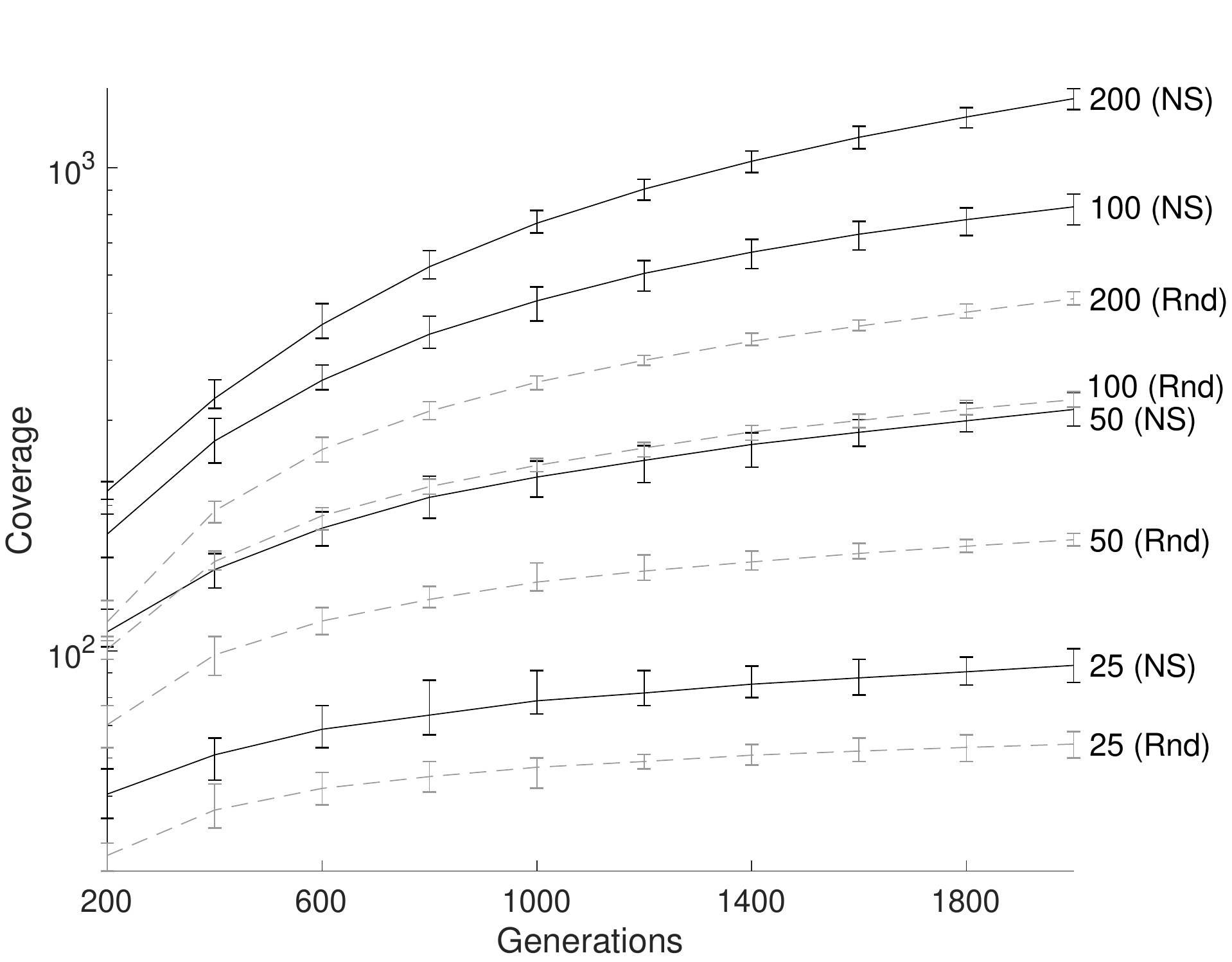}\label{fig: 5x5x5 voxel}} 
\subfloat[$V_{size}=10\times10\times10$]{\includegraphics[width=0.3\textwidth,trim=1.5cm 7.5cm 0cm 7cm,clip]{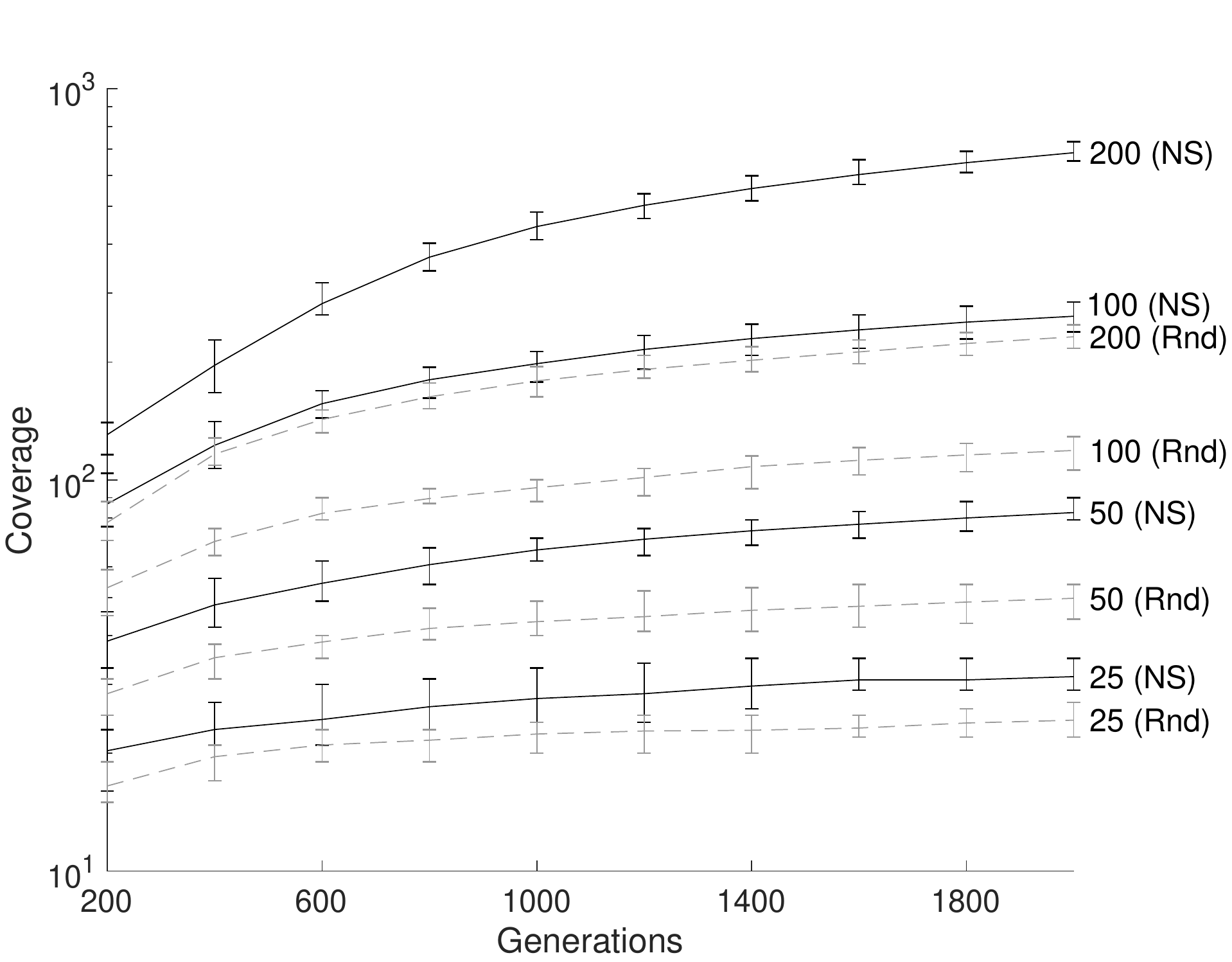}\label{fig: 10x10x10 voxel}} 

\subfloat[Compared behaviour spaces]{\includegraphics[width=0.9\textwidth,trim=2.9cm 6.2cm 3cm 7cm,clip]{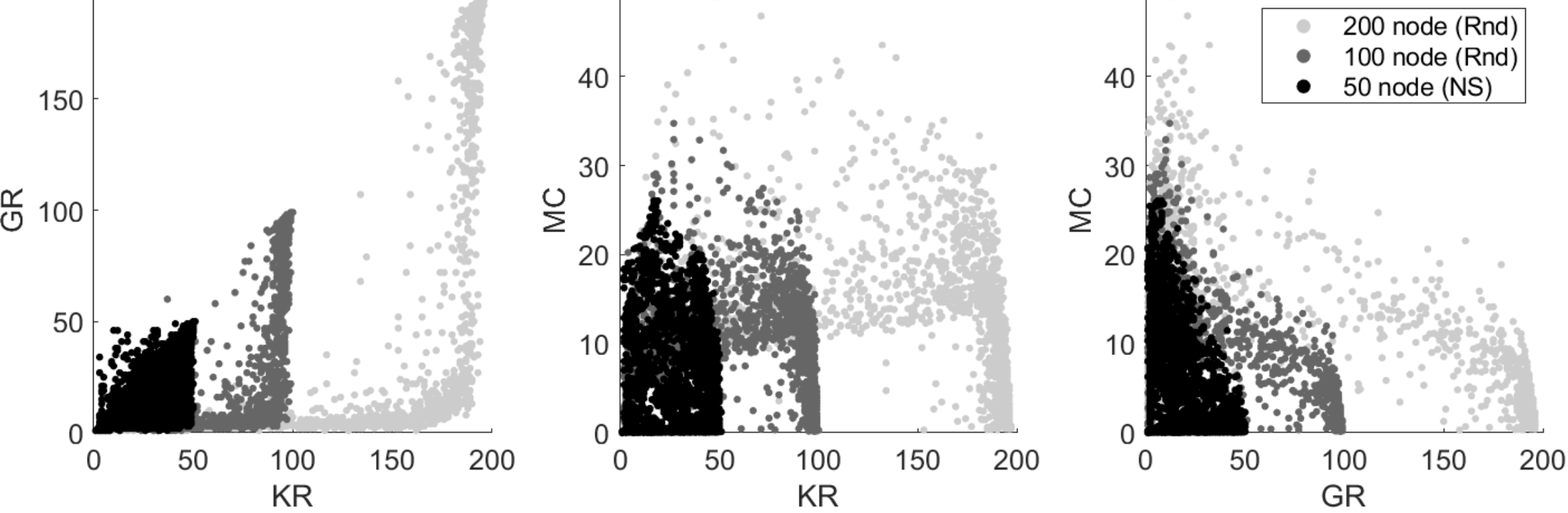}\label{fig: ESN BS overlay}} 

\caption{Average coverage (over 10 runs) of behaviour space against number of generations. Four network sizes and the two search methods are shown: Novelty Search (black, solid line) and random search (grey, dashed line). Error bars show minimum and maximum coverage. Quality is given for all network sizes, for both random and novelty search. \label{fig: voxel measures} }
\end{figure*}

The coverages of the different search methods and network sizes using the minimal voxel size are given in Fig. \ref{fig: 1x1x1 voxel}. Here we see that small voxel sizes significantly overestimate the dynamical freedom/quality of networks explored using novelty search. Smaller networks such as the 25 and 50 node ESNs are seen to occupy similar or more voxels than larger networks, suggesting similar or better quality. However, when visually comparing each explored behaviour space (see Fig.~\ref{fig: ESN BS overlay}) we see the measure fails to account for diversity and spread of behaviours. This demonstrates the importance of selecting an appropriate voxel size, as a voxel size too small can not differentiate between local areas that are highly populated, and fewer behaviours spread across a greater distance. 

To reduce the problem, the voxel size must be increased. By how much depends on the size of the behaviour spaces being compared. As a guide, when comparing drastically different systems, a larger voxel size will tend to differentiate better. Of course, a voxel size too big will also struggle to differentiate between systems. Because of this potential biasing problem, a visual inspection of the behaviour space is always recommended. Examples of different voxel sizes are given in Fig.~\ref{fig: voxel measures}. 

% In Fig.~\ref{fig: 10x10x10 voxel}, the coverage of each network size and search method is plotted as the number of generations increase. Here, we see the large voxel size differentiates fairly well between different network sizes. Novelty Search is again shown to explore further than random search as the number of evaluations increases. 

%coverage: NS vs random voxel 1x1x1
%%%%%%%%%%%%%%%%%%%%%%%%%%% Figure %%%%%%%%%%%%%%%%%%%%%%%%%%%
%%%%%%%%%%%%%%%%%%%%%%%%%%% Figure %%%%%%%%%%%%%%%%%%%%%%%%%%%

\section{Reliability of Behaviour Space Representation} \label{app: accuracy BS}

%%%%%%%%%%%%%%%%%%%%%%%%%%% Figure %%%%%%%%%%%%%%%%%%%%%%%%%%%
\begin{figure}[tp]
\centering
\includegraphics[width=0.5\textwidth,trim=2.5cm 1cm 3.5cm 1cm,clip]{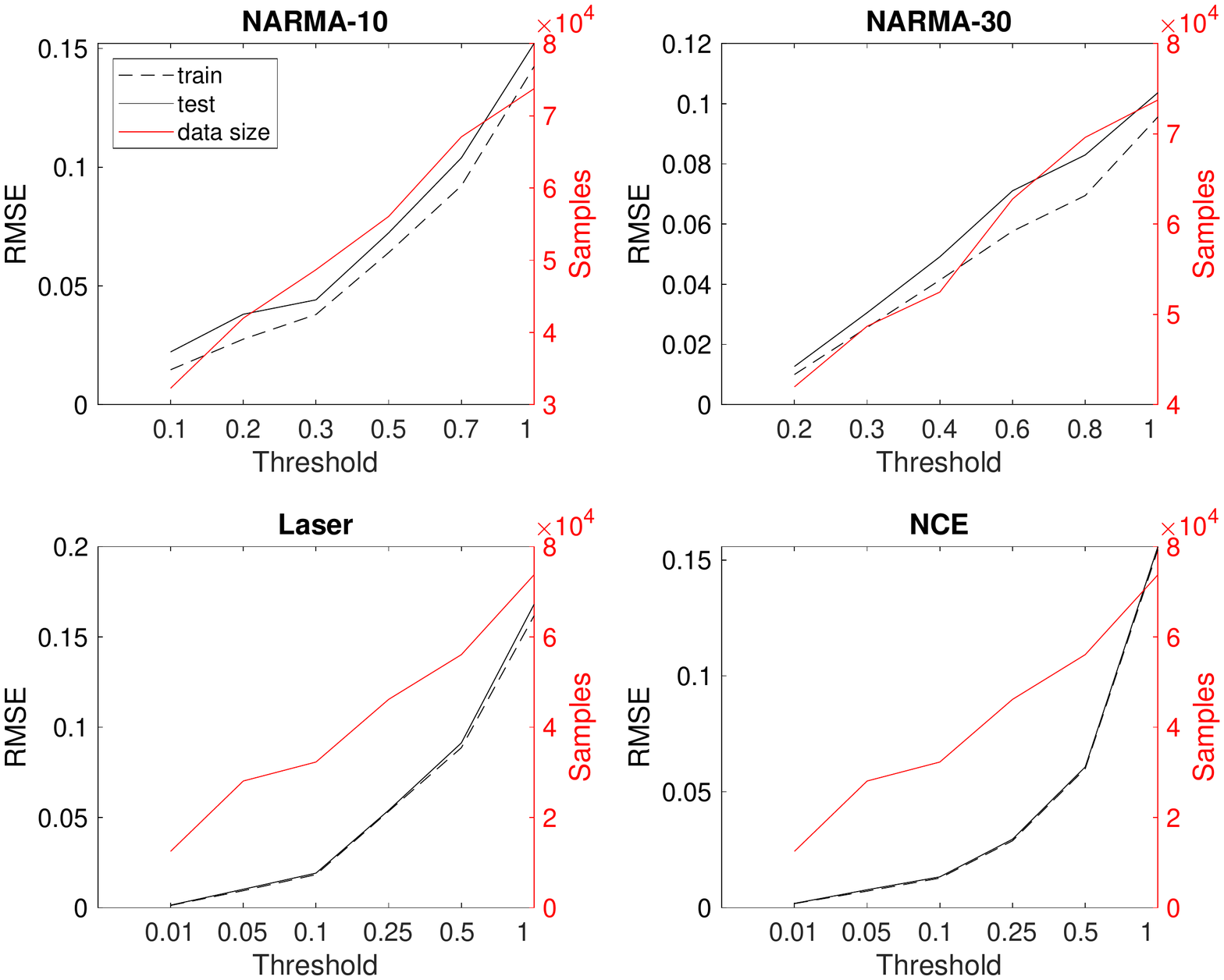}
\caption{Test accuracy (RMSE) of the FNN when trained and tested on task performances below a threshold. The number of behaviours producing task performances below the threshold is also given.\label{fig: accuracy when thresholding} }
\end{figure}
%%%%%%%%%%%%%%%%%%%%%%%%%%% Figure %%%%%%%%%%%%%%%%%%%%%%%%%%%

%%%%%%%%%%%%%%%%%%%%%%%%%%% Figure %%%%%%%%%%%%%%%%%%%%%%%%%%%
\begin{figure}[tp]
\centering
\includegraphics[width=0.5\textwidth,trim=1.5cm 7.5cm 1.5cm 8cm,clip]{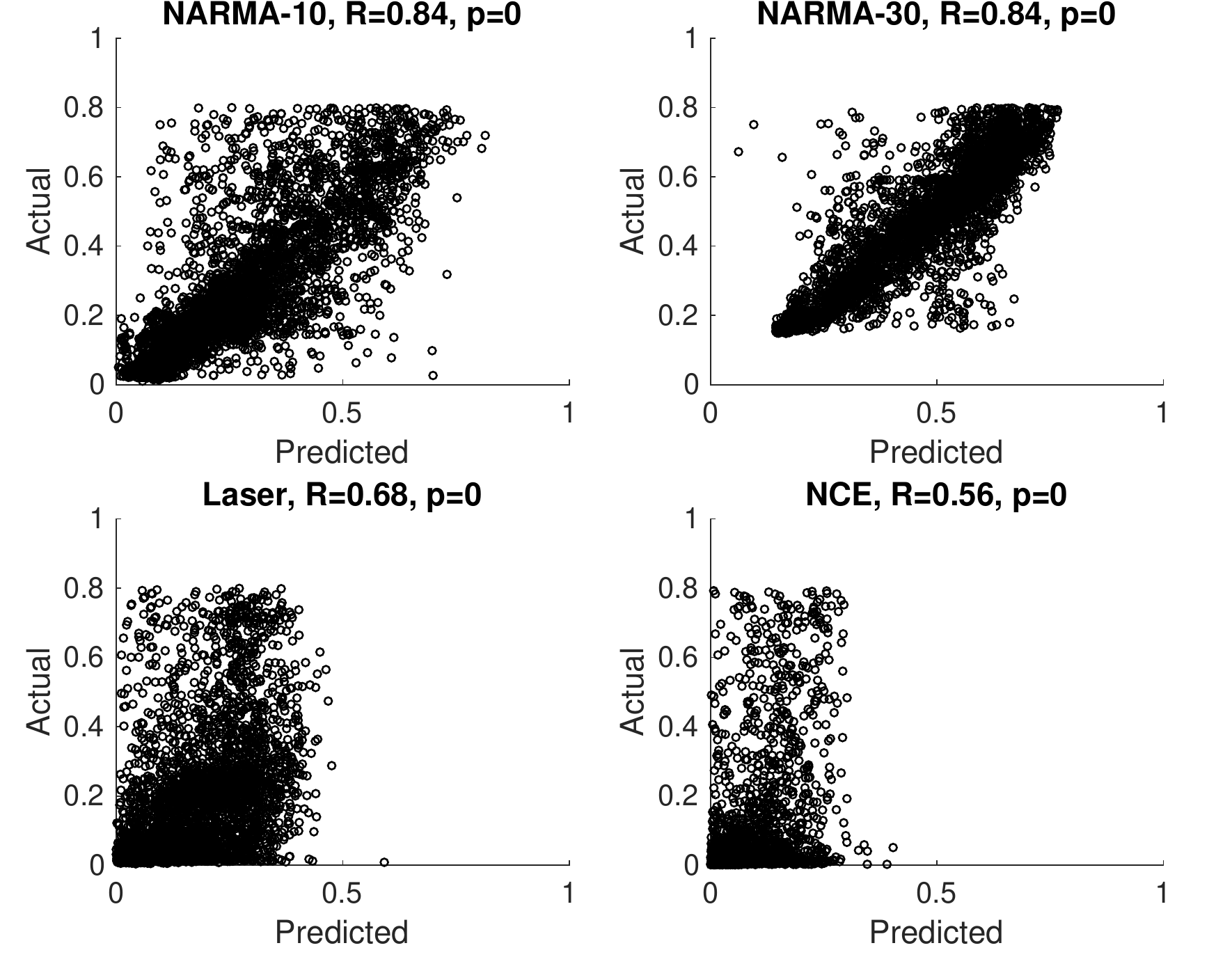}
\caption{Predicted performance versus actual performance. $\sim$7,000 task performances (NMSEs) shown for 200-node ESN. $R$ is Spearman's Rho correlation coefficient: all correlations are significant at the 95\% confidence level.
\label{fig: actual vs predicted} }
\end{figure}
%%%%%%%%%%%%%%%%%%%%%%%%%%% Figure %%%%%%%%%%%%%%%%%%%%%%%%%%%

To train the FFNNs in section~\ref{sec:phaseone}\ref{sec: accuracy of BS}, the Levenberg-Marquardt algorithm~\cite{levenberg1944method} was used for 1000 epochs, with training data set as 70\% of the database $D$, and 30\% set aside for testing. To gather statistics, 10 FFNNs were trained and tested for each network size and task.

Determining the reliability of the behaviour space representation is challenging. Selecting suitable data for this modelling process is difficult as some behaviours perform particularly poorly on tasks, reducing the overall prediction accuracy of the model. Poor task performing reservoirs, tend to increase noise in the models training data as some appear to be randomly scattered across the behaviour space. %This also suggests the behaviour representation is incomplete.

To reduce the problem, different thresholds were placed on the training data to show how well the relationship of the highest performing reservoirs can be modelled. Applying each threshold,  reservoirs with task performance (NMSE) above the threshold are removed from the training and test data.

A low prediction error (RMSE) of the model with low thresholds, indicates greater ability to predict high performing reservoirs. At higher thresholds, more training data is available but includes reservoirs that perform poorly on the task.   

The mean prediction error of 10 feed-forward networks, trained with each threshold, on each task, using the behaviours from the 200 node ESNs, are shown in Fig.\ref{fig: accuracy when thresholding}. Across all tasks, the accuracy of the model improves when smaller thresholds are used, i.e., error is smallest when predicting only the highest performing reservoirs, suggesting a strong relationship between behaviour space and task performance.

To visualise how well the relationship is modelled for task performances of NMSE$<1$, we plot the predicted NMSE versus the evaluated NMSE in Fig.\ref{fig: actual vs predicted}. Here, the output of four FFNNs, trained for each task, are given. We see the laser and non-linear channel equalisation tasks are harder to model, typically resulting in an overestimation, as the actual task performances of most behaviours tend to be low, generally with an NMSE $<0.2$. 
We also calculate Spearman's Rho (called $R$ here), a non-parametric test measuring the strength of association between predicted and actual. A value of $-1$ indicates perfect negative correlation, while a value of $+1$ indicates perfect positive correlation. A value of $0$ indicates no correlation between predicted and actual.
The $p$ value for each measure is also provided. If the $p$ value is less than the significance level of 0.05, it indicates a rejection of the null hypothesis that no correlation exists, at the 95\% confidence level. The high values of $R$ and essentially zero $p$ values suggest the models predict task performance very well.

\end{document}